\documentclass[prd,two column,aps,floats,floatfix,preprintnumbers,amssymb,nofootinbib,longbibliography]{revtex4-1}
\usepackage{graphicx}
\usepackage{microtype,graphicx}
\usepackage[english]{babel}
\let\oldfootnote\footnote
\renewcommand{\footnote}[1]{%
    \begingroup%
    \linespread{1}
    \oldfootnote{#1}%
    \endgroup%
}
\usepackage{amsthm}
\usepackage{enumitem}
\usepackage{latexsym}
\usepackage{amsfonts}
\usepackage{amssymb}
\usepackage{caption}
\usepackage[dvipsnames]{xcolor}
\usepackage{CJK}
\usepackage[export]{adjustbox}
\usepackage{amsmath}
\usepackage{slashed}
\usepackage{dcolumn}
\usepackage{verbatim}
\usepackage{appendix}
\usepackage{float}
\usepackage{multirow}
\usepackage{soul}
\usepackage[normalem]{ulem}
\usepackage{ulem}
\usepackage{hyperref}
\usepackage{adjustbox}
\usepackage{natbib}
\usepackage{epsfig}
\usepackage{nicefrac}

  \theoremstyle{plain}

\usepackage[english]{babel}
  \providecommand{\definitionname}{Definition}
  \providecommand{\propositionname}{Proposition}
  \providecommand{\theoremname}{Theorem}
  \providecommand{\corollaryname}{Corollary}
  \providecommand{\conjecturename}{Conjecture}

\def\Mpl{M_{pl}}
\def\mpl{M_{\rm Pl}}

\def\eg{{\em e.g.\/}}

\def\beq{\begin{equation}}
\def\eeq{\end{equation}}
\def\be{\begin{equation}}
\def\ee{\end{equation}}
\def\bea{\begin{eqnarray}}
\def\eea{\end{eqnarray}}

\newcommand{\eq}[1]{Eq.~\ref{#1}}

\def\m{\mu}

\begin{document}
\count\footins = 1000 

\title{Nonsingular black holes as dark matter}

\author{Paul~C.~W.~Davies 
\thanks{paul.davies@asu.edu}}
\affiliation{ 
Department of Physics \& Beyond Center for Fundamental Concepts in Science,  
Arizona State University, Tempe, AZ 85287-1504, USA}
\author{Damien A. Easson\thanks{easson@asu.edu}}
\affiliation{ 
Department of Physics \& Beyond Center for Fundamental Concepts in Science,  
Arizona State University, Tempe, AZ 85287-1504, USA}
\author{Phillip B.~Levin\thanks{philliplevin@asu.edu}}
\affiliation{ 
Department of Physics \& Beyond Center for Fundamental Concepts in Science,  
Arizona State University, Tempe, AZ 85287-1504, USA}

\date{\today}

\begin{abstract}
It is commonly assumed that low-mass primordial black holes cannot constitute a significant fraction of the dark matter in our universe due to their predicted short lifetimes from the conventional Hawking radiation and evaporation process. Assuming physical black holes are nonsingular--likely due to quantum gravity or other high-energy physics--we demonstrate that a large class of nonsingular black holes have finite evaporation temperatures. This can lead to slowly evaporating low-mass black holes or to remnant mass states that circumvent traditional evaporation constraints. As a proof of concept, we explore the limiting curvature hypothesis and the evaporation process of a nonsingular black hole solution in two-dimensional dilaton gravity. We identify generic features of the radiation profile and compare them with known regular black holes, such as the Bardeen solution in four dimensions. Remnant masses are proportional to the fundamental length scale, and we argue that slowly evaporating low-mass nonsingular black holes, or remnants, are viable dark matter candidates.
\end{abstract}
\maketitle

\section{Introduction} The story of dark matter began in 1933 with Fritz Zwicky's observation that the visible matter in the Coma cluster of galaxies was insufficient to account for its observed gravitational effects \cite{Zwicky:1933gu}. In the 1970s, Vera Rubin's studies of galactic rotation curves provided further compelling evidence \cite{Rubin:1980zd}. She discovered that stars in the outer regions of galaxies were orbiting at unexpectedly high speeds, challenging Newtonian predictions and suggesting the presence of unseen mass surrounding the galaxies. Additional support for dark matter comes from gravitational lensing, galaxy rotation curves, the cosmic microwave background and the formation of cosmic structures. Assuming general relativity, dark, non-baryonic matter is estimated to comprise nearly 25\% of the total matter content of the universe~\cite{DES:2022ccp}. Despite decades of research and significant investment in direct detection experiments—including searches for weakly interacting massive particles (WIMPs) and axions—no conclusive evidence of dark matter particles has yet been found.

Many studies have shown that primordial black holes (PBHs) can be excellent dark matter candidates (see, \eg~\cite{Belotsky:2014kca,Green:2020jor,Villanueva-Domingo:2021spv,Carr:2024nlv}).
Stephen Hawking’s discovery that black holes radiate via quantum processes near the event horizon introduced the idea that black holes are not perfectly ``black" but instead emit radiation by losing mass, leading to several important questions about what happens as they shrink in size. As the mass decreases, the temperature increases, leading to even higher-intensity radiation and eventually reaching a point where the black hole should evaporate entirely, leaving nothing behind. This is predicted to result in a culminating burst of high-energy particles as the black hole disappears.~\cite{Hawking:1974rv}.

\begin{figure}[H]
\centering
  \includegraphics[width=1\linewidth]{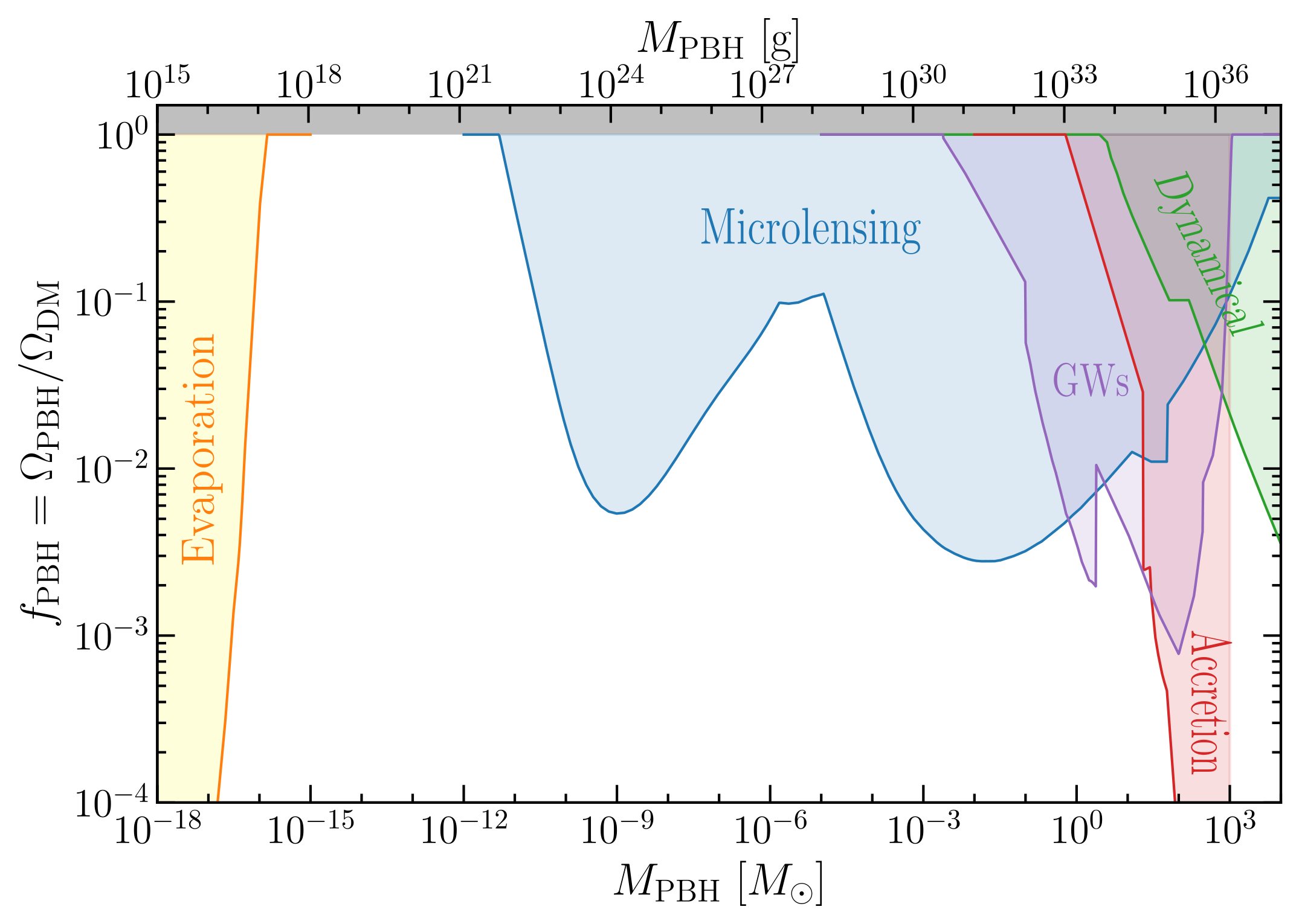}
   \captionof{figure}{Constraints on the fraction of DM in the form of PBHs, $f_{\mathrm{PBHs}}$, as a function of mass, assuming a monochromatic mass function. The bounds shown (left to right) are from evaporation (yellow), microlensing (blue), merger rates from gravitational waves (pink), accretion (red), dynamical disruption (green). For each bound, the tightest constraint at each mass is shown and the shaded regions are excluded under standard assumptions. Figure created using PlotPBHbounds \cite{bradley_j_kavanagh_2019_3538999}, which is regularly updated to include the latest constraints.}
  \label{PBHFig}
\end{figure}

Conventional calculations for a black hole of mass $m$ lead to a Hawking temperature, $T_H \propto 1/m$, with luminosity, $P \propto 1/m^2$.  
A solar mass black hole takes more than $10^{67}$ years to evaporate--significantly longer than the current age of the universe ($1.38 \times 10^{10}$~years), while for a black hole of $10^{11}$~kg, the evaporation time is $2.6 \times10^9$~years. Thus, experiments are searching for signs of exploding primordial black holes. To date, no such radiation has been observed. The evaporation observations place tight constraints on the viability of low-mass PBHs as dark matter, see yellow region of Fig.~\ref{PBHFig}. 

As the mass of the black hole approaches zero, the temperature approaches infinity, indicating the semi-classical description of black hole evaporation via Hawking radiation eventually breaks down and quantum gravity effects or other new physics are expected to play a significant role,
potentially altering the final stages of black hole evaporation. 

The standard evaporation process leads to another key issue: the black hole information loss problem \cite{Polchinski:2016hrw}. Consistent quantum field theories indicate that information must be preserved, and yet once the black hole fully evaporates, there seems to be no trace of this information, leading to a direct conflict with the principles of quantum mechanics.
Thus, our current understanding of black holes is incomplete, as it leads to a violation of unitarity—a cornerstone of quantum theory.

In this paper, we demonstrate that the infinity in the Hawking temperature is inextricably linked to another famous infinity: the singularity at the center of the black hole interior. 
We show how nonsingular black holes can naturally evolve to form stable remnants after the evaporation process.
Indeed, early proposals that black hole evaporation resulted in Planck mass remnants argued that remnants were necessary for unitary time evolution, and hypothesized that these remnants could be dark matter \cite{Aharonov:1987tp,MacGibbon:1987my}. 
Many others followed \cite{Barrow:1992hq,PhysRevD.46.645,Adler:2001vs,Easson:2002tg,Chen:2002tu,Cai:2010zh,Dymnikova:2015yma,Carr:2021bzv,Profumo:2024fxq, Calza:2024fzo, Calza:2024xdh}.

We argue that the natural evaporation process for nonsingular black holes reopens the evaporation window shown in Fig.~\ref{PBHFig}, extending it by many orders of magnitude to the left, potentially down to the Planck mass scale, $\Mpl \sim 10^{-8} \, \text{kg}$. This allows nonsingular, low-mass PBHs to be viable candidates for dark matter, either as remnants or slowly-evaporating black holes.

\section{2D dilaton gravity model}
To explore our hypothesis, consider a simplified and tractable 2D model. In the context of $(1+1)$ dilaton gravity, one may construct a nonsingular Schwarzschild-de Sitter (SdS) black hole metric \cite{Easson:2017pfe} by utilizing an extremal curvature conjecture, which combines the maximal curvature conjecture of \cite{Markov1982} with the minimal curvature conjecture of \cite{Easson:2006jd}. 
The metric for this dilaton gravity black hole (DGBH) may be written:
\begin{equation}
    \mathrm{d}s^2 = -n(r) \mathrm{d}t^2 + n(r)^{-1} \mathrm{d}r^2 \,,
\end{equation}
with $n(r)$ given by:
\begin{widetext}
\begin{multline}\label{eqg00}
     n(r) = \frac{1}{3}\left(\frac{m}{l}\right)^{\nicefrac{2}{3}}\left(1+ \frac{l^2 \Lambda}{3}\right)\ln{\frac{r^2-\left(ml^2\right)^{\nicefrac{1}{3}}r+\left(ml^2\right)^{\nicefrac{2}{3}}}{(r + \left(ml^2\right)^{\nicefrac{1}{3}})^2}}  \\ +
    \frac{2}{\sqrt{3}}\left(\frac{m}{l}\right)^{\nicefrac{2}{3}}\left(1 + \frac{l^2 \Lambda}{3}\right)\arctan{\left(\frac{2r - (ml^2)^{\nicefrac{1}{3}}}{\sqrt{3}(ml^2)^{\nicefrac{1}{3}}}\right)} - \frac{\Lambda}{3}r^2,
\end{multline}
\end{widetext}
where $m$ is the mass of the black hole, $l$ is the fundamental length scale, and $\Lambda$ is the cosmological constant. 
%
This spacetime describes a singularity-free black hole in a de Sitter background. The solution has two horizons: a black hole horizon and a cosmological horizon. An observer falling through the black hole horizon approaches a smooth, constant, maximally curved spacetime, with Ricci scalar, $R_{max} \propto l^{-2}$. Outside of this horizon the solution resembles the familiar Schwarzschild-de Sitter solution and particle detectors respond to the thermal effects of both horizons. 

\section{Hawking radiation of nonsingular black holes}
For our current investigation we consider \eq{eqg00}, with $\Lambda= 0$. 
To study the Hawking radiation process, we use the trace anomaly to obtain the flux of energy due to Hawking radiation \cite{PhysRevD.15.2088}. The trace  of the anomalous energy-momentum tensor $T^{\mu}_{\mu}$ may be written in terms of curvature invariants and several cases in 2D are analyzed in \cite{Drehmer2009}. In \( (1 + 1) \) dimensions, the anomaly is expressed solely in terms of the Ricci scalar \( R \), the only independent curvature invariant in 2D \cite{Deser:1976yx}:
\begin{equation}
    T^{\mu}_{\mu} = \frac{R}{24\pi} \,.
\end{equation}
For our metric, $R = -n''(r)$, where $\prime \equiv d/dr$,  so that:
\begin{equation}
    T_\m{}^\m = \frac{1}{24\pi} \frac{2m(2r^3-ml^2)}{(r^3 + ml^2)^2}.
\end{equation}
The conservation equations for our time-independent $T^{\nu}_{\mu}$ are:
\begin{equation}
    \partial_r T_r^r = 0 \,, \qquad \partial_r T^r_r = \frac{n'}{2 n} (T^t_t - T^r_r) \,,
\end{equation}
the latter of which is equivalent to 
\be
\partial_r ( n \,T_r^r )=  \frac{n'}{2} T_\m^\m
\,.
\ee

The full stress-energy tensor is given by:
\begin{eqnarray}
    T^{*\beta}_{\alpha} &=&
    \begin{pmatrix}
        T_\m{}^\m(r)  - n^{-1}(r) H_{2D}(r) & 0 \\
        0 & n^{-1}(r) H_{2D}(r)
    \end{pmatrix} \nonumber \\
    &+& \frac{A}{n(r)}
    \begin{pmatrix}
        1 & -1 \\
        1 & -1
    \end{pmatrix}
    + \frac{B}{n(r)}
    \begin{pmatrix}
        -1 & 0 \\
        0 & -1
    \end{pmatrix}\,,
\end{eqnarray} 
where $A$ and $B$ are integration constants and $H_{2D}(r)$ is defined as
\begin{eqnarray}
    H_{2D}(r) &=& \int_{r_{0}}^{r} \frac{n'(r')}{2}T^{\mu}_{\mu}(r') \,dr' \\ \nonumber
    &=& \frac{-m^2}{24 \pi}\left(\frac{r^2}{(ml^2 + r^3)^2} - \frac{r_{0}^2}{(ml^2 + r_{0}^3)^2}\right) \,.
\end{eqnarray}
To determine $A$ and $B$,  transform the stress-energy tensor into null coordinates, $u = t + r^*$ and $v = t - r^*$, where $r^*$ is the tortoise coordinate:
\begin{eqnarray}
   T_{uu} &=& \frac{1}{4} \left[ 2B + 2H_{2D}(r) - n(r) T^\m_\m(r) \right ]\,, \nonumber \\ 
   T_{uv} &=& T_{vu} = - \frac{1}{4}n(r) T^\m_\m(r) \,, \nonumber \\
   T_{vv} &=& T_{uu} - A \,.
\end{eqnarray}
Hawking radiation is the quantum vacuum state satisfying the Unruh vacuum boundary conditions. There is no flux of ingoing particles at infinity, $T_{vv}(r \rightarrow \infty) =0$, and no energy flux at the past horizon, $T_{uu}(V \rightarrow 0) = 0$, giving
\be
A = \lim_{r\to\infty} \left(T_{uu}(r) - T_{vv}(r)\right) \,, \qquad B=0
\,.
\ee
Therefore, the energy flux of particles created is:
\begin{equation}
    T^{r}_{t}(r \rightarrow \infty)=A=\frac{\kappa^2}{48\pi} + \frac{1}{48\pi} \lim_{r \rightarrow \infty}\left[n(r)n''(r) - \frac{n'(r)^2}{4}\right]
\end{equation}
For our metric, the above limit vanishes, so the flux depends only on the surface gravity $\kappa$:
\begin{equation}\label{eqflux}
     \frac{\kappa^2}{48\pi} = \frac{m^2 r_{0}^2}{48\pi (r_{0}^3 + ml^2)^2} \,,
\end{equation}
The corresponding Hawking temperature is proportional to $\kappa$:
\begin{equation}
    T_{H} = \frac{\kappa}{2 \pi} = \frac{mr_{0}}{2\pi(r_{0}^3 + ml^2)}.
\end{equation}
This result is in agreement with previous findings using the method of complex paths \cite{Easson:2002tg}.
Unlike the Schwarzschild Hawking temperature,  this temperature reaches a maximum and then the black hole begins to cool, possibly settling as a remnant mass (see DGBH, Fig.~\ref{fig1}). The flux, \eq{eqflux},
is equal to  $dm/dt$ so that,
\begin{equation}
t = 48 \pi \int \frac{(r_{0}^3 + m' l^2)^2}{m'^2 r_{0}^2} \, dm' \,.
\end{equation}
We are unable to analytically find an exact solution for the horizon position $r_0$ and compute the above integral;  
however, towards the late stages of the black hole evaporation we consider the limit as $r \rightarrow 0$ to estimate the remnant mass.
If a remnant is ultimately formed, its mass $m_{*}$ can be calculated by expanding $n(r)$ and solving for the mass in the limit that the horizon vanishes. 
Series expanding $n(r)$ to second order yields:
\begin{equation}
    n(r) = 1 - \frac{4 \pi}{3 \sqrt{3}}\left(\frac{m}{l}\right)^{\nicefrac{2}{3}} + \frac{r^2}{l^2} \,,
\end{equation}
where we have introduced an integration constant from the asymptotically flat requirement $n \rightarrow 1$ as $r \rightarrow \infty$.
Setting the above approximation to zero gives the horizon location $r_0$:
\begin{equation}
    r_{0} = \frac{l}{3} \sqrt{4 \sqrt{3} \pi \left(\frac{m}{l}\right)^{\nicefrac{2}{3}} - 9}\,,
\end{equation}
from which the remnant black hole mass $m_{*}$ is calculated in the limit $r_{0} \rightarrow 0$:
\begin{equation}
    m_\star = \frac{9}{8} \sqrt{\frac{\sqrt{3}}{\pi^3}} \,l \,.
 \label{remnantMass}
\end{equation}
As expected, the remnant mass is proportional to the minimum length scale, $l$. Conventional MKS units are restored by multiplying the mass by $\frac{c^2}{G}$.
Setting $l$ to the Planck length, $l_p = 1.6 \times 10^{-35}$ $m$, yields,
\begin{equation}
    m_{*}(l_{p}) = 5.7 \times 10^{-9} \ kg = 3.2 \times 10^{18} \ GeV/c^{2}.
\end{equation}
This value rests well within the range for dark matter candidates between  $10^{-22} eV$ -- $10^{48} GeV$, making the remnant neither suspiciously light nor heavy~\cite{Rajendran:2022kcs}. 

\section{4D nonsingular black holes}

Having established our proof of concept, we now move to four dimensional models.
We consider spherically symmetric metrics:
\begin{equation}\label{gmunus}
    \mathrm{d}s^2 = -f(r) \mathrm{d}t^2 + f(r)^{-1} \mathrm{d}r^2 +  r^2 \left(\mathrm{d}\theta^2 + \sin^2 \mathrm{d}\phi^2 \right),
\end{equation}
with
\begin{equation}\label{mofr}
    f(r) = 1 - \frac{2 M(r)}{r} \,,
\end{equation}
where $M(r) = m$ for Schwarzschild.
The 4D nonsingular blackholes we study are in Table~\ref{table1}, constructed in \cite{bardeen1968proceedings,Hayward:2005gi,Fan:2016hvf,Dymnikova:1992ux}.

\begin{table}[h!]
\centering
\begin{tabular}{|>{\centering\arraybackslash}m{3cm}|>{\centering\arraybackslash}m{5cm}|}
    \hline
    \textbf{BH Solution} & \textbf{Mass Function $M(r)$} \\
    \hline
    Bardeen & $\frac{m r^3}{(r^2 + l^2)^{3/2}}$ \\
    \hline
    Hayward & $\frac{m r^3}{r^3 + 2 m l^2}$ \\
    \hline
    Fan-Wang & $\frac{m r^3}{(r + l)^3}$ \\
    \hline
    Dymnikova & $\frac{2m}{\pi} \left( \arctan\left( \frac{r}{l} \right) - \frac{l r}{r^2 + l^2} \right)$ \\
    \hline
\end{tabular} 
\caption{Mass functions for regular black hole models.}
\label{table1}
\end{table}

\begin{figure}[H]
\centering
\includegraphics[width=1\linewidth]{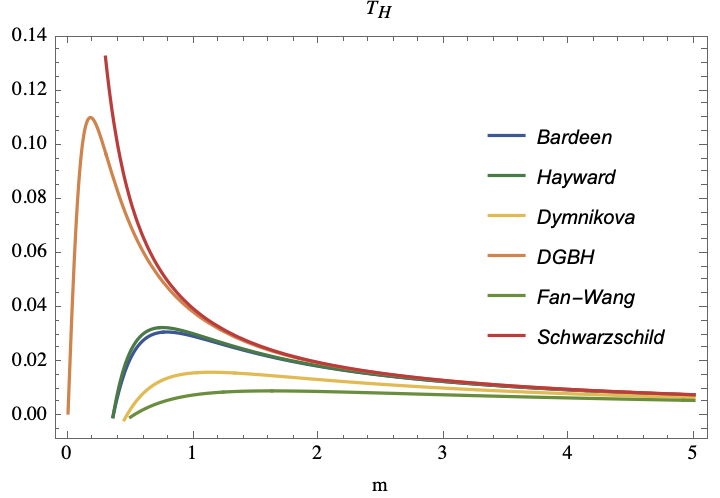}
\caption{Hawking temperature $T_H$ ploted as a function of mass $m$. While the temperature of a singular black hole increases without bound as the mass decreases (red curve), for nonsingular black holes the temperature initially rises, reaches a maximum, and then gradually decreases as the mass continues to diminish, asymptotically approaching zero. This leads to a slowly radiating black hole or ultimately a remnant mass.
}
\label{fig1}
\end{figure}

The spacetimes are asymptotically flat and $m$ is the ADM mass. The Hawking temperatures may be calculated in terms of the surface gravity $\kappa$, as illustrated in the 2D case above. We observe that the qualitative features of nonsingular models exhibit a common trend: the black hole temperature reaches a maximum and then gradually falls to zero, likely leaving behind a remnant mass as in the 2D DGBH case (see Fig.~\ref{fig1}).\footnote{Yet another way such remnants can form in quantum gravity arises in the context of asymptotically safe gravity \cite{Cai:2010zh}.}.
\section{New window for PBH dark matter}
We propose that dark matter may consist of nonsingular black holes and their asymptotic remnants formed in the primordial universe. These PBHs could have originated after inflation, during radiation domination, and before Big-Bang nucleosynthesis, from the collapse of large density perturbations and may constitute a significant fraction of the current matter-energy density if they formed as massive objects on the order of solar to multi-solar masses. Alternatively, if smaller, asteroid-sized PBHs formed in the range of \(10^{17} \, \text{kg}\) to \(10^{21} \, \text{kg}\), they could account for all of the observed dark matter \cite{Green:2024bam}. A reasonable range for our fundamental length scale extends from the Planck scale to distances on the order of $10^{-19} \, \text{m}$, the latter corresponding to the TeV energy scale currently probed by the Large Hadron Collider (LHC). Consequently, for our black hole remnants, $10^{-35} \, \text{m} \leq l \leq 10^{-19} \, \text{m}$.

\subsection{Dark Matter Abundance of Remnants}
Next, we estimate whether these remnants could constitute the dark matter of the universe. Primordial black holes can form during the early universe from the collapse of overdense regions. Such density fluctuations generally have the form:
\begin{equation}
    \delta = \epsilon \left(\frac{m}{m_{0}}\right)^{-n},
\end{equation}
where $m_{0}$ is the initial mass within the cosmological horizon at the moment of formation, $0 \leq \epsilon \leq 1$ and $n \geq 0$.

The primordial black hole mass spectrum is the number density of black holes of mass $m$ formed from the collapse of the overdense regions~\cite{Carr:1975qj}:
\begin{equation}
    n(m) = \mu_{0} F m_{0}^{-2} \epsilon \exp\left({-\frac{\beta^4}{2 \epsilon^2}}\right)\left(\frac{m}{m_{0}}\right)^{-\frac{2(1+2w)}{1+w}} , \label{numberDensity}
\end{equation}
where $\mu_{0}$ is the density of the universe at the time of primordial black hole formation, $F$ is the ratio of the number density today to number density initially, $\beta$ is the fractional collapse, and $w$ is the equation of state parameter relating the fluid pressure $p$ to the energy density $\rho$, via $p = w \rho$. Assuming that the primordial black hole remnants form shortly after the end of inflation, $w = 1/3$.

Integrating Eq. \eqref{numberDensity} gives the initial mass density of primordial black holes:
\begin{equation}
    \rho_{PBH}^{initial} = 2 \mu_{0} F \epsilon \exp\left({\frac{-\beta^4}{2 \epsilon^2}}\right)\left[\sqrt{\frac{m_{0}}{m_{lower}}} - \sqrt{\frac{m_{0}}{m_{upper}}}\right],
\end{equation}
where $m_{upper}$ and $m_{lower}$ are the largest and smallest masses of the primordial black holes formed. 
For $m_{upper}>\!\!>m_{lower}$, 
$\rho_{PBH}^{initial} \approx 1/\sqrt{m_{lower}}$.
The fraction of dark matter today is $\Omega_{PBH}^{today} = a^{-3} \Omega_{PBH}^{initial}$, and the fraction of dark matter in primordial remnants is $\Omega_{PBH}^{today} \approx a^{-3}/\sqrt{m_{lower}}$. 

The traditional argument asserts that primordial black holes with masses smaller than $10^{15} \, \text{g}$ would have evaporated by now due to Hawking radiation, thus setting the lower bound of the mass spectrum at $m_{\text{lower}} \approx 10^{15} \, \text{g}$. For slightly larger PBHs, their Hawking radiation would be detectable in the extragalactic gamma-ray background, as well as in the \( e^\pm \) and antiproton fluxes. The absence of these fluxes places constraints on the abundance of light PBHs~\cite{MacGibbon:1991vc,Carr:2009jm,Laha:2019ssq,Boudaud:2018hqb}.

Our nonsingular black holes do not fully evaporate. This significantly extends the lower mass limit, allowing $m_{\text{lower}} \approx l$. By choosing $m_{\text{lower}} \approx m_{\text{Planck}}$, corresponding to $l \approx 10^{-35} \, \text{m}$, and $m_{\text{upper}} \approx 10^{22} \, \text{kg}$, we find $\Omega_{\text{PBH}}^{\text{today}} \approx 0.23$, with the parameters from Eq.~\eqref{numberDensity} falling within the expected range \cite{Carr:1975qj}. Therefore, near Planck-mass nonsingular black holes or remnants could constitute all of the observed dark matter today, evading observational constraints set by conventional radiation calculations.

\section{Conclusions}
In addition to being dark matter candidates, nonsingular black holes have the potential to address other deep mysteries, such as the information loss paradox \cite{Polchinski:2016hrw}. They may provide baby universe interiors, or some other place to store in-falling information \cite{Easson:2001qf}. This work should be viewed as a foundational and admittedly speculative exploration. The results are primarily theoretical but potentially profound. The stability and observational characteristics of these remnants depend on the existence of a minimal length scale, and the effective field theories that prevent singularity formation, which likely violate traditional energy conditions~\cite{Hawking:1970zqf,ZASLAVSKII2010278,Maeda:2021jdc} (see Appendix \ref{ecnsbh}). Future advances in quantum gravity and observational techniques will be essential to test the robustness of nonsingular black holes as dark matter candidates. 

\acknowledgments
We thank M.~Baumgart, B.~Carr, M.~Parikh, B.~Ratra and S.~Vagnozzi for valuable discussions and correspondence. DAE is supported in part by the U.S. Department of Energy, Office of High Energy Physics, under Award Number DE-SC0019470.

\appendix
\section{Properties of supporting matter}\label{ecnsbh}
The Birkhoff theorem establishes that the singular Schwarzschild black hole metric is the unique spherically symmetric vacuum solution to the Einstein field equations \cite{Birkhoff1923}—a profound result that also extends to Kleinian (split-signature) metrics \cite{Easson:2023ytf}. The theorem ensures that any nonsingular spherically symmetric metric, including those studied here, must be supported by matter in standard Einstein gravity. 

Natural questions arise such as, what is the origin of this supporting matter which ultimately must violate traditional energy conditions to prevent singularity formation? How does this matter interact with ordinary matter from the Standard Model? Ultimately the interactions of this unusual supporting matter must be strongly constrained to avoid giving rise to long range ‘fifth forces’ and violations of the equivalence principle. The theories must obey solar system tests of gravity. While the properties of this supporting matter are exotic, at the classical level the alternative to this matter is formation of a singularity in gravitational collapse, which is arguably the most pathological option in a physical theory. In this appendix, we explore the nature of this supporting matter, beginning with an analysis of the energy conditions it must satisfy or violate.~\footnote{We do not analyze the energy conditions for the metric \eq{eqg00}, as the model is formulated in 2D dilaton gravity, where known subtleties arise in defining the energy conditions \cite{Chatterjee:2012zh}.}. 

\subsection{Energy Conditions}
It is well known that the traditional energy conditions must be violated by matter in order to avoid singularities \cite{Penrose:1964wq,Hawking:1966sx,Hawking:1966jv,Hawking:1967ju,Hawking:1970zqf}. The theorems identify the necessary criteria for establishing geodesic completeness in both stationary and time-dependent cosmological spacetimes \cite{Lesnefsky:2022fen,Easson:2024uxe,Easson:2024fzn}. We now examine the degree of the energy condition violation needed to support the nonsingular solutions of Table I. 

The regular black holes are supported by effective energy momentum tensors of Type I \cite{S.W.Hawking1973}, whose canonical form in the local Lorentz frame is 
\begin{equation}
    T^{(\hat{a})(\hat{b})}  =  T^{\hat\mu\hat{\nu}}e^{(\hat{a})}_{\hat\mu}e^{(\hat{b})}_{\hat\nu} \,,
\end{equation}
where $\{ e_{(\hat{a})}^{\hat{\mu}} \}$ with $a = 0,1,2,3$, is a set of orthonormal basis vectors in the local Lorentz frame having 
$e_{(\hat a)}^{\hat{\nu}}e_{(\hat b) \hat\nu}= \eta_{(\hat{a})(\hat{b})}$. Here $\eta_{(\hat{a})(\hat{b})}$ is a Minkowski metric in the local Lorentz frame
and the curved metric is  $g_{\hat{\mu}\hat{\nu}}=\eta_{(\hat{a})(\hat{b})}e^{(\hat{a})}_{\hat{\mu}}e^{(\hat{b})}_{\hat{\nu}}$.
Energy density and pressure are given by the compentns of the mixed Einstein tensor:
\begin{equation}
    T^{\hat{\mu}}{}_{\hat{\nu}} = G^{\hat{\mu}}{}_{\hat{\nu}} = diag(-\rho, p_{\parallel}, p_\perp, p_\perp) \,,
\end{equation}
where we have taken the reduced Planck mass $\mpl^{-2} = 8 \pi G =1$. For the models considered here the energy density is $\rho = - T^{\hat{t}}{}_{\hat{t}}$ and we have pressures parallel to the radial coordinate, $p_\parallel = T^{\hat{r}}{}_{\hat{r}}$ and perpendicular to the radial coordinate, $p_\perp = T^{\hat{\theta}}{}_{\hat{\theta}}= T^{\hat{\phi}}{}_{\hat{\phi}}$. It is a further well-know property of our spacetimes that, $T^{\hat{r}}{}_{\hat{r}}= T^{\hat{t}}{}_{\hat{t}}$.

For the metric \eq{gmunus} with $M(r)$ in \eq{mofr}, 
\begin{equation}
\rho = - p_\parallel = \frac{2 M'(r)}{r^2}\,, \qquad p_\perp = -\frac{M''(r)}{r} \,.
\end{equation}

We analyze the standard energy conditions.  The null energy condition (NEC) requires both $\rho +p_\parallel  \geq 0$  and $\rho +p_\perp  \geq 0$ for all values of the parameters.  The weak energy condition (WEC) requires $\rho \ge 0$ and $\rho +p_\perp  \ge 0$ and $\rho +p_\parallel  \ge 0$. The dominant energy condition (DEC): $\rho \ge 0$ , $p_\perp \in [-\rho,\rho]$ and $p_\parallel \in [-\rho,\rho]$. The strong energy condition (SEC): $\rho + p_\parallel + 2p_\perp \geq0$. 
We see, DEC$\implies$WEC, WEC$\implies$NEC, SEC$\implies$NEC; and SEC$\rlap{$\quad\not$}\implies$WEC.
\begin{figure}[H]
\centering
  \includegraphics[width=1\linewidth]{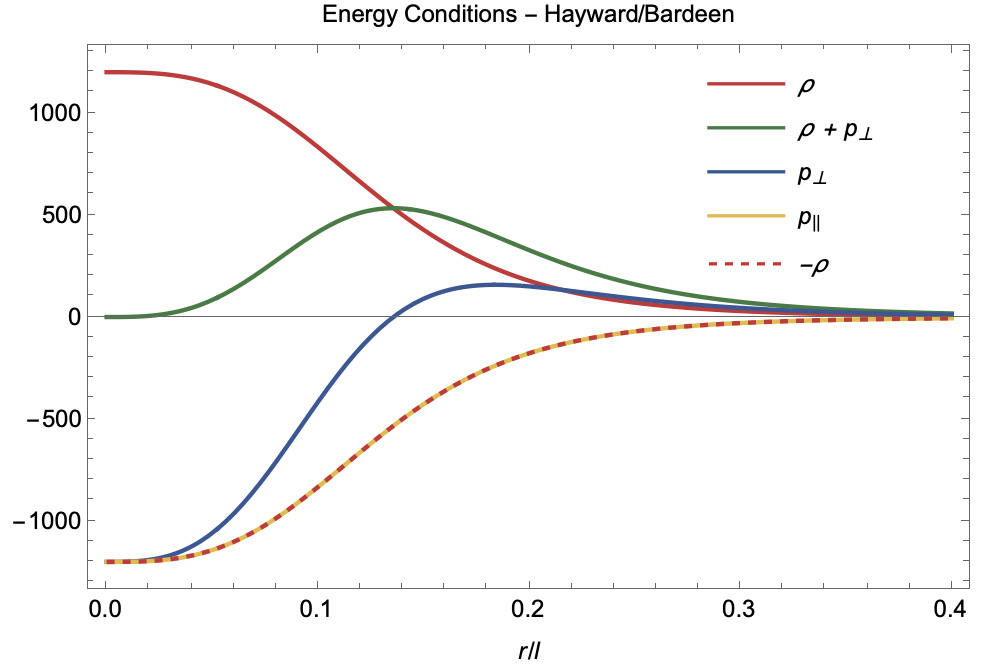}
   \captionof{figure}{Energy condition plot for the Hayward spacetime which is qualitatively similar to Bardeen. Legend quantities are rendered dimensionless by the factors $M^2_{pl}$ and $l$.  Model parameters: $m= 1$, $l=.5$.}
  \label{ecfig}
\end{figure}

\begin{figure}[H]
\centering
  \includegraphics[width=1\linewidth]{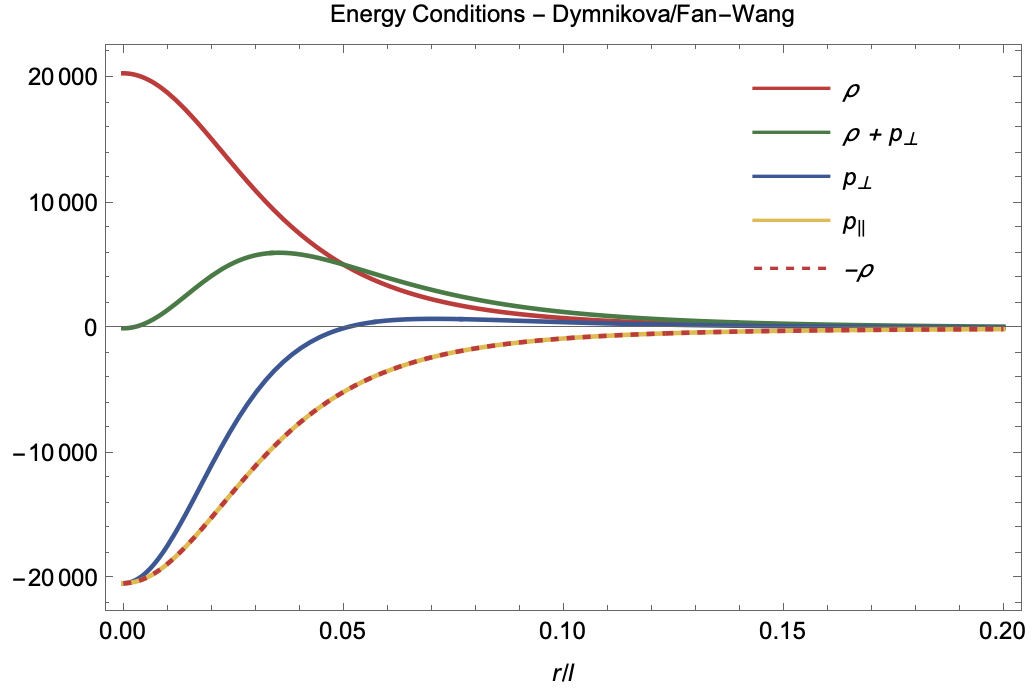}
   \captionof{figure}{Energy condition plot for the Dymnikova spacetime which is qualitatively similar to Fan-Wang. Legend quantities are rendered dimensionless by the factors $M^2_{pl}$ and $l$.  Model parameters: $m= 1$, $l=.05$.}
  \label{ec2fig}
\end{figure}

In most of our models, coordinates span the ranges $t \in (-\infty, \, \infty)$, $r \in (-\infty, \, \infty)$, $\theta \in [ 0, \, \pi ]$ and $\phi \in (-\pi,\, \pi]$;  For our purposes of studying energy conditions it is sufficient to focus our plots in the range $r \in [0, \, \infty)$; however, some of spacetimes such as Bardeen cover the full range $r \in (-\infty, \, \infty)$ (see, Fig.~9 of \cite{Easson:2020esh}.) For physically reasonable values of parameters resulting in positive mass and black hole horizons, we find qualitatively similar energy condition behaviors for Bardeen and Hayward spacetimes (see Fig.~\ref{ecfig}) and similarly for Dymnikova and Fan-Wang spacetimes (see Fig.~\ref{ec2fig}).

From Fig.~\ref{ecfig}, (for the given parameters) we see the Bardeen and Hayward spacetimes obey the NEC and WEC for the entire domain of $r$, as depicted by the positivity of the solid red and green curves and the equality of the dotted red and yellow curves, $p_\parallel = - \rho$. The DEC is violated in the asymptotically flat region as seen from the blue curve superseding the red curve at large $r$. The SEC is violated (as expected) near the de Sitter core as seen by the negativity of the blue curve near $r=0$. 

From Fig.~\ref{ec2fig}, (for the given parameters) we see the Dymnikova and Fan-Wang spacetimes obey the NEC, WEC and DEC for the entire domain of $r$, as depicted by the positivity of the solid red and green curves and the equality of the dotted red and yellow curves, $p_\parallel = - \rho$. The DEC is satisfied as the blue curve remains bound by the red and dotted red curves for all $r$. Only the SEC is violated near the de Sitter core for $r \leq l$.

\subsection{Matter Lagrangian}
After evaluating the energy conditions that must be violated to sustain the regular black hole geometries in Table I, we now briefly discuss the type of matter Lagrangian required to source these spacetimes and how such spacetimes may form from gravitational collapse. 

In \cite{Ayon-Beato:1998hmi,Ayon-Beato:2000mjt} it was shown that the Bardeen regular black hole spacetime can be interpreted as an exact solution in general relativity with a source arising from nonlinear electrodynamics (NLED). A key characteristic of NLED models is that gauge field quantities remain finite everywhere, including at the location of a point source, inspiring the authors to explore the Bardeen black hole as a solution sourced by a monopole within an NLED framework.

The source is a magnetic monopole with charge $g$, with a Lagrangian given by
\begin{equation}\label{NLEDlagrangian}
\mathcal{L}=\frac{3m}{|g|^3}\Bigg(\frac{\sqrt{2g^2F}}{1+\sqrt{2g^2F}}\Bigg)^{5/2},
\end{equation}
where $F\equiv\frac{1}{4}F_{\mu\nu}F^{\mu\nu}$ and $m$ is the black hole mass. The ansatz for the gauge field and metric 
\begin{equation}\label{monopolemet}
-g_{tt}(r)=1-\frac{2M(r)}{r}, \ \ \ \ \ \ \ \ F_{\mu\nu}=2\delta^\theta_{[\mu}\delta^\phi_{\nu]}B(r,\theta),
\end{equation}
become solutions to the Einstein-NLED equations when
\begin{equation}
M(r)=\frac{mr^3}{(r^2+g^2)^{3/2}}, \ \ \ \ \ \ \ B(\theta)=g\sin\theta.
\end{equation}
This form matches the Bardeen black hole with $g\rightarrow l$.

The other models of Table I may follow from similar NLED constructions.  As there are several ways a black hole can be made regular using matter, the Birkoff theorem no longer applies and one may be concerned that the end state of general gravitational spherical collapse may no longer be unique. Might there be a physical reason to favor one regular black hole metric over another? It is intriguing that the Bardeen solution appears to emerge from fundamental principles in string theory such as T-duality \cite{Gaete:2022ukm}.

\subsection{Formation and observational signatures}
Recently, in \cite{Bueno:2024eig,Bueno:2024zsx} it is argued that dynamical gravitational collapse in a theory with an infinite tower of higher-derivative corrections to the Einstein-Hilbert action, under very general conditions, leads to the formation of regular black holes.

Finally, advances in the formation process of regular black holes has led to the exciting possibility of observational signatures of regular black holes, including features in regular black hole shadows from phase transitions during collapse potentially delaying apparent horizon formation \cite{Yan:2019hxx,Balart:2024rtj,Vertogradov:2025snh}.

\bibliography{physics}

\begin{thebibliography}{64}%
\makeatletter
\providecommand \@ifxundefined [1]{%
 \@ifx{#1\undefined}
}%
\providecommand \@ifnum [1]{%
 \ifnum #1\expandafter \@firstoftwo
 \else \expandafter \@secondoftwo
 \fi
}%
\providecommand \@ifx [1]{%
 \ifx #1\expandafter \@firstoftwo
 \else \expandafter \@secondoftwo
 \fi
}%
\providecommand \natexlab [1]{#1}%
\providecommand \enquote  [1]{``#1''}%
\providecommand \bibnamefont  [1]{#1}%
\providecommand \bibfnamefont [1]{#1}%
\providecommand \citenamefont [1]{#1}%
\providecommand \href@noop [0]{\@secondoftwo}%
\providecommand \href [0]{\begingroup \@sanitize@url \@href}%
\providecommand \@href[1]{\@@startlink{#1}\@@href}%
\providecommand \@@href[1]{\endgroup#1\@@endlink}%
\providecommand \@sanitize@url [0]{\catcode `\\12\catcode `\$12\catcode `\&12\catcode `\#12\catcode `\^12\catcode `\_12\catcode `\%12\relax}%
\providecommand \@@startlink[1]{}%
\providecommand \@@endlink[0]{}%
\providecommand \url  [0]{\begingroup\@sanitize@url \@url }%
\providecommand \@url [1]{\endgroup\@href {#1}{\urlprefix }}%
\providecommand \urlprefix  [0]{URL }%
\providecommand \Eprint [0]{\href }%
\providecommand \doibase [0]{http://dx.doi.org/}%
\providecommand \selectlanguage [0]{\@gobble}%
\providecommand \bibinfo  [0]{\@secondoftwo}%
\providecommand \bibfield  [0]{\@secondoftwo}%
\providecommand \translation [1]{[#1]}%
\providecommand \BibitemOpen [0]{}%
\providecommand \bibitemStop [0]{}%
\providecommand \bibitemNoStop [0]{.\EOS\space}%
\providecommand \EOS [0]{\spacefactor3000\relax}%
\providecommand \BibitemShut  [1]{\csname bibitem#1\endcsname}%
\let\auto@bib@innerbib\@empty
\bibitem [{\citenamefont {Zwicky}(1933)}]{Zwicky:1933gu}%
  \BibitemOpen
  \bibfield  {author} {\bibinfo {author} {\bibfnamefont {F.}~\bibnamefont {Zwicky}},\ }\bibfield  {title} {\enquote {\bibinfo {title} {{Die Rotverschiebung von extragalaktischen Nebeln}},}\ }\href {\doibase 10.1007/s10714-008-0707-4} {\bibfield  {journal} {\bibinfo  {journal} {Helv. Phys. Acta}\ }\textbf {\bibinfo {volume} {6}},\ \bibinfo {pages} {110--127} (\bibinfo {year} {1933})}\BibitemShut {NoStop}%
\bibitem [{\citenamefont {Rubin}\ \emph {et~al.}(1980)\citenamefont {Rubin}, \citenamefont {Thonnard},\ and\ \citenamefont {Ford}}]{Rubin:1980zd}%
  \BibitemOpen
  \bibfield  {author} {\bibinfo {author} {\bibfnamefont {V.~C.}\ \bibnamefont {Rubin}}, \bibinfo {author} {\bibfnamefont {N.}~\bibnamefont {Thonnard}}, \ and\ \bibinfo {author} {\bibfnamefont {W.~K.}\ \bibnamefont {Ford}, \bibfnamefont {Jr.}},\ }\bibfield  {title} {\enquote {\bibinfo {title} {{Rotational properties of 21 SC galaxies with a large range of luminosities and radii, from NGC 4605 /R = 4kpc/ to UGC 2885 /R = 122 kpc/}},}\ }\href {\doibase 10.1086/158003} {\bibfield  {journal} {\bibinfo  {journal} {Astrophys. J.}\ }\textbf {\bibinfo {volume} {238}},\ \bibinfo {pages} {471} (\bibinfo {year} {1980})}\BibitemShut {NoStop}%
\bibitem [{\citenamefont {Abbott}\ \emph {et~al.}(2023)\citenamefont {Abbott} \emph {et~al.}}]{DES:2022ccp}%
  \BibitemOpen
  \bibfield  {author} {\bibinfo {author} {\bibfnamefont {T.~M.~C.}\ \bibnamefont {Abbott}} \emph {et~al.} (\bibinfo {collaboration} {DES}),\ }\bibfield  {title} {\enquote {\bibinfo {title} {{Dark Energy Survey Year 3 results: Constraints on extensions to \ensuremath{\Lambda}CDM with weak lensing and galaxy clustering}},}\ }\href {\doibase 10.1103/PhysRevD.107.083504} {\bibfield  {journal} {\bibinfo  {journal} {Phys. Rev. D}\ }\textbf {\bibinfo {volume} {107}},\ \bibinfo {pages} {083504} (\bibinfo {year} {2023})},\ \Eprint {http://arxiv.org/abs/2207.05766} {arXiv:2207.05766 [astro-ph.CO]} \BibitemShut {NoStop}%
\bibitem [{\citenamefont {Belotsky}\ \emph {et~al.}(2014)\citenamefont {Belotsky}, \citenamefont {Dmitriev}, \citenamefont {Esipova}, \citenamefont {Gani}, \citenamefont {Grobov}, \citenamefont {Khlopov}, \citenamefont {Kirillov}, \citenamefont {Rubin},\ and\ \citenamefont {Svadkovsky}}]{Belotsky:2014kca}%
  \BibitemOpen
  \bibfield  {author} {\bibinfo {author} {\bibfnamefont {K.~M.}\ \bibnamefont {Belotsky}}, \bibinfo {author} {\bibfnamefont {A.~D.}\ \bibnamefont {Dmitriev}}, \bibinfo {author} {\bibfnamefont {E.~A.}\ \bibnamefont {Esipova}}, \bibinfo {author} {\bibfnamefont {V.~A.}\ \bibnamefont {Gani}}, \bibinfo {author} {\bibfnamefont {A.~V.}\ \bibnamefont {Grobov}}, \bibinfo {author} {\bibfnamefont {M.~Yu.}\ \bibnamefont {Khlopov}}, \bibinfo {author} {\bibfnamefont {A.~A.}\ \bibnamefont {Kirillov}}, \bibinfo {author} {\bibfnamefont {S.~G.}\ \bibnamefont {Rubin}}, \ and\ \bibinfo {author} {\bibfnamefont {I.~V.}\ \bibnamefont {Svadkovsky}},\ }\bibfield  {title} {\enquote {\bibinfo {title} {{Signatures of primordial black hole dark matter}},}\ }\href {\doibase 10.1142/S0217732314400057} {\bibfield  {journal} {\bibinfo  {journal} {Mod. Phys. Lett. A}\ }\textbf {\bibinfo {volume} {29}},\ \bibinfo {pages} {1440005} (\bibinfo {year} {2014})},\ \Eprint {http://arxiv.org/abs/1410.0203} {arXiv:1410.0203 [astro-ph.CO]} \BibitemShut
  {NoStop}%
\bibitem [{\citenamefont {Green}\ and\ \citenamefont {Kavanagh}(2021)}]{Green:2020jor}%
  \BibitemOpen
  \bibfield  {author} {\bibinfo {author} {\bibfnamefont {Anne~M.}\ \bibnamefont {Green}}\ and\ \bibinfo {author} {\bibfnamefont {Bradley~J.}\ \bibnamefont {Kavanagh}},\ }\bibfield  {title} {\enquote {\bibinfo {title} {{Primordial Black Holes as a dark matter candidate}},}\ }\href {\doibase 10.1088/1361-6471/abc534} {\bibfield  {journal} {\bibinfo  {journal} {J. Phys. G}\ }\textbf {\bibinfo {volume} {48}},\ \bibinfo {pages} {043001} (\bibinfo {year} {2021})},\ \Eprint {http://arxiv.org/abs/2007.10722} {arXiv:2007.10722 [astro-ph.CO]} \BibitemShut {NoStop}%
\bibitem [{\citenamefont {Villanueva-Domingo}\ \emph {et~al.}(2021)\citenamefont {Villanueva-Domingo}, \citenamefont {Mena},\ and\ \citenamefont {Palomares-Ruiz}}]{Villanueva-Domingo:2021spv}%
  \BibitemOpen
  \bibfield  {author} {\bibinfo {author} {\bibfnamefont {Pablo}\ \bibnamefont {Villanueva-Domingo}}, \bibinfo {author} {\bibfnamefont {Olga}\ \bibnamefont {Mena}}, \ and\ \bibinfo {author} {\bibfnamefont {Sergio}\ \bibnamefont {Palomares-Ruiz}},\ }\bibfield  {title} {\enquote {\bibinfo {title} {{A brief review on primordial black holes as dark matter}},}\ }\href {\doibase 10.3389/fspas.2021.681084} {\bibfield  {journal} {\bibinfo  {journal} {Front. Astron. Space Sci.}\ }\textbf {\bibinfo {volume} {8}},\ \bibinfo {pages} {87} (\bibinfo {year} {2021})},\ \Eprint {http://arxiv.org/abs/2103.12087} {arXiv:2103.12087 [astro-ph.CO]} \BibitemShut {NoStop}%
\bibitem [{\citenamefont {Carr}\ and\ \citenamefont {Green}(2024)}]{Carr:2024nlv}%
  \BibitemOpen
  \bibfield  {author} {\bibinfo {author} {\bibfnamefont {Bernard~J.}\ \bibnamefont {Carr}}\ and\ \bibinfo {author} {\bibfnamefont {Anne~M.}\ \bibnamefont {Green}},\ }\bibfield  {title} {\enquote {\bibinfo {title} {{The History of Primordial Black Holes}},}\ }\href@noop {} {\  (\bibinfo {year} {2024})},\ \Eprint {http://arxiv.org/abs/2406.05736} {arXiv:2406.05736 [astro-ph.CO]} \BibitemShut {NoStop}%
\bibitem [{\citenamefont {Hawking}(1974)}]{Hawking:1974rv}%
  \BibitemOpen
  \bibfield  {author} {\bibinfo {author} {\bibfnamefont {S.~W.}\ \bibnamefont {Hawking}},\ }\bibfield  {title} {\enquote {\bibinfo {title} {{Black hole explosions}},}\ }\href {\doibase 10.1038/248030a0} {\bibfield  {journal} {\bibinfo  {journal} {Nature}\ }\textbf {\bibinfo {volume} {248}},\ \bibinfo {pages} {30--31} (\bibinfo {year} {1974})}\BibitemShut {NoStop}%
\bibitem [{\citenamefont {Kavanagh}(2019)}]{bradley_j_kavanagh_2019_3538999}%
  \BibitemOpen
  \bibfield  {author} {\bibinfo {author} {\bibfnamefont {Bradley~J.}\ \bibnamefont {Kavanagh}},\ }\href {\doibase 10.5281/zenodo.3538999} {\enquote {\bibinfo {title} {bradkav/pbhbounds: Release version},}\ } (\bibinfo {year} {2019})\BibitemShut {NoStop}%
\bibitem [{\citenamefont {Polchinski}(2017)}]{Polchinski:2016hrw}%
  \BibitemOpen
  \bibfield  {author} {\bibinfo {author} {\bibfnamefont {Joseph}\ \bibnamefont {Polchinski}},\ }\bibfield  {title} {\enquote {\bibinfo {title} {{The black hole information problem.}}}\ }in\ \href {\doibase 10.1142/9789813149441_0006} {\emph {\bibinfo {booktitle} {{Theoretical Advanced Study Institute in Elementary Particle Physics}: {New Frontiers in Fields and Strings}}}}\ (\bibinfo {year} {2017})\ pp.\ \bibinfo {pages} {353--397},\ \Eprint {http://arxiv.org/abs/1609.04036} {arXiv:1609.04036 [hep-th]} \BibitemShut {NoStop}%
\bibitem [{\citenamefont {Aharonov}\ \emph {et~al.}(1987)\citenamefont {Aharonov}, \citenamefont {Casher},\ and\ \citenamefont {Nussinov}}]{Aharonov:1987tp}%
  \BibitemOpen
  \bibfield  {author} {\bibinfo {author} {\bibfnamefont {Y.}~\bibnamefont {Aharonov}}, \bibinfo {author} {\bibfnamefont {A.}~\bibnamefont {Casher}}, \ and\ \bibinfo {author} {\bibfnamefont {S.}~\bibnamefont {Nussinov}},\ }\bibfield  {title} {\enquote {\bibinfo {title} {{The Unitarity Puzzle and Planck Mass Stable Particles}},}\ }\href {\doibase 10.1016/0370-2693(87)91320-7} {\bibfield  {journal} {\bibinfo  {journal} {Phys. Lett. B}\ }\textbf {\bibinfo {volume} {191}},\ \bibinfo {pages} {51} (\bibinfo {year} {1987})}\BibitemShut {NoStop}%
\bibitem [{\citenamefont {MacGibbon}(1987)}]{MacGibbon:1987my}%
  \BibitemOpen
  \bibfield  {author} {\bibinfo {author} {\bibfnamefont {Jane~H.}\ \bibnamefont {MacGibbon}},\ }\bibfield  {title} {\enquote {\bibinfo {title} {{Can Planck-mass relics of evaporating black holes close the universe?}}}\ }\href {\doibase 10.1038/329308a0} {\bibfield  {journal} {\bibinfo  {journal} {Nature}\ }\textbf {\bibinfo {volume} {329}},\ \bibinfo {pages} {308--309} (\bibinfo {year} {1987})}\BibitemShut {NoStop}%
\bibitem [{\citenamefont {Barrow}(1992)}]{Barrow:1992hq}%
  \BibitemOpen
  \bibfield  {author} {\bibinfo {author} {\bibfnamefont {J.~D.}\ \bibnamefont {Barrow}},\ }\bibfield  {title} {\enquote {\bibinfo {title} {{Thermodynamics of open universes}},}\ }\href {\doibase 10.1103/PhysRevD.46.R3227} {\bibfield  {journal} {\bibinfo  {journal} {Phys. Rev. D}\ }\textbf {\bibinfo {volume} {46}},\ \bibinfo {pages} {R3227--R3230} (\bibinfo {year} {1992})}\BibitemShut {NoStop}%
\bibitem [{\citenamefont {Barrow}\ \emph {et~al.}(1992)\citenamefont {Barrow}, \citenamefont {Copeland},\ and\ \citenamefont {Liddle}}]{PhysRevD.46.645}%
  \BibitemOpen
  \bibfield  {author} {\bibinfo {author} {\bibfnamefont {John~D.}\ \bibnamefont {Barrow}}, \bibinfo {author} {\bibfnamefont {Edmund~J.}\ \bibnamefont {Copeland}}, \ and\ \bibinfo {author} {\bibfnamefont {Andrew~R.}\ \bibnamefont {Liddle}},\ }\bibfield  {title} {\enquote {\bibinfo {title} {The cosmology of black hole relics},}\ }\href {\doibase 10.1103/PhysRevD.46.645} {\bibfield  {journal} {\bibinfo  {journal} {Phys. Rev. D}\ }\textbf {\bibinfo {volume} {46}},\ \bibinfo {pages} {645--657} (\bibinfo {year} {1992})}\BibitemShut {NoStop}%
\bibitem [{\citenamefont {Adler}\ \emph {et~al.}(2001)\citenamefont {Adler}, \citenamefont {Chen},\ and\ \citenamefont {Santiago}}]{Adler:2001vs}%
  \BibitemOpen
  \bibfield  {author} {\bibinfo {author} {\bibfnamefont {Ronald~J.}\ \bibnamefont {Adler}}, \bibinfo {author} {\bibfnamefont {Pisin}\ \bibnamefont {Chen}}, \ and\ \bibinfo {author} {\bibfnamefont {David~I.}\ \bibnamefont {Santiago}},\ }\bibfield  {title} {\enquote {\bibinfo {title} {{The Generalized uncertainty principle and black hole remnants}},}\ }\href {\doibase 10.1023/A:1015281430411} {\bibfield  {journal} {\bibinfo  {journal} {Gen. Rel. Grav.}\ }\textbf {\bibinfo {volume} {33}},\ \bibinfo {pages} {2101--2108} (\bibinfo {year} {2001})},\ \Eprint {http://arxiv.org/abs/gr-qc/0106080} {arXiv:gr-qc/0106080} \BibitemShut {NoStop}%
\bibitem [{\citenamefont {Easson}(2003)}]{Easson:2002tg}%
  \BibitemOpen
  \bibfield  {author} {\bibinfo {author} {\bibfnamefont {Damien~A.}\ \bibnamefont {Easson}},\ }\bibfield  {title} {\enquote {\bibinfo {title} {{Hawking radiation of nonsingular black holes in two-dimensions}},}\ }\href {\doibase 10.1088/1126-6708/2003/02/037} {\bibfield  {journal} {\bibinfo  {journal} {JHEP}\ }\textbf {\bibinfo {volume} {02}},\ \bibinfo {pages} {037} (\bibinfo {year} {2003})},\ \Eprint {http://arxiv.org/abs/hep-th/0210016} {arXiv:hep-th/0210016} \BibitemShut {NoStop}%
\bibitem [{\citenamefont {Chen}\ and\ \citenamefont {Adler}(2003)}]{Chen:2002tu}%
  \BibitemOpen
  \bibfield  {author} {\bibinfo {author} {\bibfnamefont {Pisin}\ \bibnamefont {Chen}}\ and\ \bibinfo {author} {\bibfnamefont {Ronald~J.}\ \bibnamefont {Adler}},\ }\bibfield  {title} {\enquote {\bibinfo {title} {{Black hole remnants and dark matter}},}\ }\href {\doibase 10.1016/S0920-5632(03)02088-7} {\bibfield  {journal} {\bibinfo  {journal} {Nucl. Phys. B Proc. Suppl.}\ }\textbf {\bibinfo {volume} {124}},\ \bibinfo {pages} {103--106} (\bibinfo {year} {2003})},\ \Eprint {http://arxiv.org/abs/gr-qc/0205106} {arXiv:gr-qc/0205106} \BibitemShut {NoStop}%
\bibitem [{\citenamefont {Cai}\ and\ \citenamefont {Easson}(2010)}]{Cai:2010zh}%
  \BibitemOpen
  \bibfield  {author} {\bibinfo {author} {\bibfnamefont {Yi-Fu}\ \bibnamefont {Cai}}\ and\ \bibinfo {author} {\bibfnamefont {Damien~A.}\ \bibnamefont {Easson}},\ }\bibfield  {title} {\enquote {\bibinfo {title} {{Black holes in an asymptotically safe gravity theory with higher derivatives}},}\ }\href {\doibase 10.1088/1475-7516/2010/09/002} {\bibfield  {journal} {\bibinfo  {journal} {JCAP}\ }\textbf {\bibinfo {volume} {09}},\ \bibinfo {pages} {002} (\bibinfo {year} {2010})},\ \Eprint {http://arxiv.org/abs/1007.1317} {arXiv:1007.1317 [hep-th]} \BibitemShut {NoStop}%
\bibitem [{\citenamefont {Dymnikova}\ and\ \citenamefont {Khlopov}(2015)}]{Dymnikova:2015yma}%
  \BibitemOpen
  \bibfield  {author} {\bibinfo {author} {\bibfnamefont {Irina}\ \bibnamefont {Dymnikova}}\ and\ \bibinfo {author} {\bibfnamefont {Maxim}\ \bibnamefont {Khlopov}},\ }\bibfield  {title} {\enquote {\bibinfo {title} {{Regular black hole remnants and graviatoms with de Sitter interior as heavy dark matter candidates probing inhomogeneity of early universe}},}\ }\href {\doibase 10.1142/S0218271815450029} {\bibfield  {journal} {\bibinfo  {journal} {Int. J. Mod. Phys. D}\ }\textbf {\bibinfo {volume} {24}},\ \bibinfo {pages} {1545002} (\bibinfo {year} {2015})},\ \Eprint {http://arxiv.org/abs/1510.01351} {arXiv:1510.01351 [gr-qc]} \BibitemShut {NoStop}%
\bibitem [{\citenamefont {Carr}\ and\ \citenamefont {Kuhnel}(2022)}]{Carr:2021bzv}%
  \BibitemOpen
  \bibfield  {author} {\bibinfo {author} {\bibfnamefont {Bernard}\ \bibnamefont {Carr}}\ and\ \bibinfo {author} {\bibfnamefont {Florian}\ \bibnamefont {Kuhnel}},\ }\bibfield  {title} {\enquote {\bibinfo {title} {{Primordial black holes as dark matter candidates}},}\ }\href {\doibase 10.21468/SciPostPhysLectNotes.48} {\bibfield  {journal} {\bibinfo  {journal} {SciPost Phys. Lect. Notes}\ }\textbf {\bibinfo {volume} {48}},\ \bibinfo {pages} {1} (\bibinfo {year} {2022})},\ \Eprint {http://arxiv.org/abs/2110.02821} {arXiv:2110.02821 [astro-ph.CO]} \BibitemShut {NoStop}%
\bibitem [{\citenamefont {Profumo}(2024)}]{Profumo:2024fxq}%
  \BibitemOpen
  \bibfield  {author} {\bibinfo {author} {\bibfnamefont {Stefano}\ \bibnamefont {Profumo}},\ }\bibfield  {title} {\enquote {\bibinfo {title} {{Ultralight Primordial Black Holes}},}\ }\href@noop {} {\  (\bibinfo {year} {2024})},\ \Eprint {http://arxiv.org/abs/2405.00546} {arXiv:2405.00546 [astro-ph.HE]} \BibitemShut {NoStop}%
\bibitem [{\citenamefont {Calz\`a}\ \emph {et~al.}(2025{\natexlab{a}})\citenamefont {Calz\`a}, \citenamefont {Pedrotti},\ and\ \citenamefont {Vagnozzi}}]{Calza:2024fzo}%
  \BibitemOpen
  \bibfield  {author} {\bibinfo {author} {\bibfnamefont {Marco}\ \bibnamefont {Calz\`a}}, \bibinfo {author} {\bibfnamefont {Davide}\ \bibnamefont {Pedrotti}}, \ and\ \bibinfo {author} {\bibfnamefont {Sunny}\ \bibnamefont {Vagnozzi}},\ }\bibfield  {title} {\enquote {\bibinfo {title} {{Primordial regular black holes as all the dark matter. I. Time-radial-symmetric metrics}},}\ }\href {\doibase 10.1103/PhysRevD.111.024009} {\bibfield  {journal} {\bibinfo  {journal} {Phys. Rev. D}\ }\textbf {\bibinfo {volume} {111}},\ \bibinfo {pages} {024009} (\bibinfo {year} {2025}{\natexlab{a}})},\ \Eprint {http://arxiv.org/abs/2409.02804} {arXiv:2409.02804 [gr-qc]} \BibitemShut {NoStop}%
\bibitem [{\citenamefont {Calz\`a}\ \emph {et~al.}(2025{\natexlab{b}})\citenamefont {Calz\`a}, \citenamefont {Pedrotti},\ and\ \citenamefont {Vagnozzi}}]{Calza:2024xdh}%
  \BibitemOpen
  \bibfield  {author} {\bibinfo {author} {\bibfnamefont {Marco}\ \bibnamefont {Calz\`a}}, \bibinfo {author} {\bibfnamefont {Davide}\ \bibnamefont {Pedrotti}}, \ and\ \bibinfo {author} {\bibfnamefont {Sunny}\ \bibnamefont {Vagnozzi}},\ }\bibfield  {title} {\enquote {\bibinfo {title} {{Primordial regular black holes as all the dark matter. II. Non-time-radial-symmetric and loop quantum gravity-inspired metrics}},}\ }\href {\doibase 10.1103/PhysRevD.111.024010} {\bibfield  {journal} {\bibinfo  {journal} {Phys. Rev. D}\ }\textbf {\bibinfo {volume} {111}},\ \bibinfo {pages} {024010} (\bibinfo {year} {2025}{\natexlab{b}})},\ \Eprint {http://arxiv.org/abs/2409.02807} {arXiv:2409.02807 [gr-qc]} \BibitemShut {NoStop}%
\bibitem [{\citenamefont {Easson}(2018)}]{Easson:2017pfe}%
  \BibitemOpen
  \bibfield  {author} {\bibinfo {author} {\bibfnamefont {Damien~A.}\ \bibnamefont {Easson}},\ }\bibfield  {title} {\enquote {\bibinfo {title} {{Nonsingular Schwarzschild\textendash{}de Sitter black hole}},}\ }\href {\doibase 10.1088/1361-6382/aae85f} {\bibfield  {journal} {\bibinfo  {journal} {Class. Quant. Grav.}\ }\textbf {\bibinfo {volume} {35}},\ \bibinfo {pages} {235005} (\bibinfo {year} {2018})},\ \Eprint {http://arxiv.org/abs/1712.09455} {arXiv:1712.09455 [hep-th]} \BibitemShut {NoStop}%
\bibitem [{\citenamefont {Markov}(1982)}]{Markov1982}%
  \BibitemOpen
  \bibfield  {author} {\bibinfo {author} {\bibfnamefont {M.A.}\ \bibnamefont {Markov}},\ }\bibfield  {title} {\enquote {\bibinfo {title} {{Maximal Curvature Conjecture}},}\ }\href@noop {} {\bibfield  {journal} {\bibinfo  {journal} {JETP Lett.}\ }\textbf {\bibinfo {volume} {36}},\ \bibinfo {pages} {265--269} (\bibinfo {year} {1982})}\BibitemShut {NoStop}%
\bibitem [{\citenamefont {Easson}(2007)}]{Easson:2006jd}%
  \BibitemOpen
  \bibfield  {author} {\bibinfo {author} {\bibfnamefont {Damien~A.}\ \bibnamefont {Easson}},\ }\bibfield  {title} {\enquote {\bibinfo {title} {{The Accelerating Universe and a Limiting Curvature Proposal}},}\ }\href {\doibase 10.1088/1475-7516/2007/02/004} {\bibfield  {journal} {\bibinfo  {journal} {JCAP}\ }\textbf {\bibinfo {volume} {02}},\ \bibinfo {pages} {004} (\bibinfo {year} {2007})},\ \Eprint {http://arxiv.org/abs/astro-ph/0608034} {arXiv:astro-ph/0608034} \BibitemShut {NoStop}%
\bibitem [{\citenamefont {Christensen}\ and\ \citenamefont {Fulling}(1977)}]{PhysRevD.15.2088}%
  \BibitemOpen
  \bibfield  {author} {\bibinfo {author} {\bibfnamefont {S.M.}\ \bibnamefont {Christensen}}\ and\ \bibinfo {author} {\bibfnamefont {S.A.}\ \bibnamefont {Fulling}},\ }\bibfield  {title} {\enquote {\bibinfo {title} {{Trace anomalies and the Hawking effect}},}\ }\href {\doibase 10.1103/PhysRevD.15.2088} {\bibfield  {journal} {\bibinfo  {journal} {Phys. Rev. D}\ }\textbf {\bibinfo {volume} {15}},\ \bibinfo {pages} {2088--2104} (\bibinfo {year} {1977})}\BibitemShut {NoStop}%
\bibitem [{\citenamefont {Drehmer}\ \emph {et~al.}(2009)\citenamefont {Drehmer} \emph {et~al.}}]{Drehmer2009}%
  \BibitemOpen
  \bibfield  {author} {\bibinfo {author} {\bibfnamefont {B.M.}\ \bibnamefont {Drehmer}} \emph {et~al.},\ }\bibfield  {title} {\enquote {\bibinfo {title} {{Energy-momentum tensors and trace anomalies in 2D quantum field theory}},}\ }\href {\doibase 10.1103/PhysRevD.79.064019} {\bibfield  {journal} {\bibinfo  {journal} {Phys. Rev. D}\ }\textbf {\bibinfo {volume} {79}},\ \bibinfo {pages} {064019} (\bibinfo {year} {2009})}\BibitemShut {NoStop}%
\bibitem [{\citenamefont {Deser}\ \emph {et~al.}(1976)\citenamefont {Deser}, \citenamefont {Duff},\ and\ \citenamefont {Isham}}]{Deser:1976yx}%
  \BibitemOpen
  \bibfield  {author} {\bibinfo {author} {\bibfnamefont {S.}~\bibnamefont {Deser}}, \bibinfo {author} {\bibfnamefont {M.J.}\ \bibnamefont {Duff}}, \ and\ \bibinfo {author} {\bibfnamefont {C.J.}\ \bibnamefont {Isham}},\ }\bibfield  {title} {\enquote {\bibinfo {title} {{Nonlocal conformal anomalies}},}\ }\href {\doibase 10.1016/0550-3213(76)90480-6} {\bibfield  {journal} {\bibinfo  {journal} {Nucl. Phys. B}\ }\textbf {\bibinfo {volume} {111}},\ \bibinfo {pages} {45--55} (\bibinfo {year} {1976})}\BibitemShut {NoStop}%
\bibitem [{\citenamefont {Rajendran}(2022)}]{Rajendran:2022kcs}%
  \BibitemOpen
  \bibfield  {author} {\bibinfo {author} {\bibfnamefont {Surjeet}\ \bibnamefont {Rajendran}},\ }\bibfield  {title} {\enquote {\bibinfo {title} {{New directions in the search for dark matter}},}\ }\href {\doibase 10.21468/SciPostPhysLectNotes.56} {\bibfield  {journal} {\bibinfo  {journal} {SciPost Phys. Lect. Notes}\ }\textbf {\bibinfo {volume} {56}},\ \bibinfo {pages} {1} (\bibinfo {year} {2022})},\ \Eprint {http://arxiv.org/abs/2204.03085} {arXiv:2204.03085 [hep-ph]} \BibitemShut {NoStop}%
\bibitem [{\citenamefont {Bardeen}(1968)}]{bardeen1968proceedings}%
  \BibitemOpen
  \bibfield  {author} {\bibinfo {author} {\bibfnamefont {James~M}\ \bibnamefont {Bardeen}},\ }\href@noop {} {\enquote {\bibinfo {title} {Proceedings of the international conference gr5},}\ } (\bibinfo {year} {1968})\BibitemShut {NoStop}%
\bibitem [{\citenamefont {Hayward}(2006)}]{Hayward:2005gi}%
  \BibitemOpen
  \bibfield  {author} {\bibinfo {author} {\bibfnamefont {Sean~A.}\ \bibnamefont {Hayward}},\ }\bibfield  {title} {\enquote {\bibinfo {title} {{Formation and evaporation of regular black holes}},}\ }\href {\doibase 10.1103/PhysRevLett.96.031103} {\bibfield  {journal} {\bibinfo  {journal} {Phys. Rev. Lett.}\ }\textbf {\bibinfo {volume} {96}},\ \bibinfo {pages} {031103} (\bibinfo {year} {2006})},\ \Eprint {http://arxiv.org/abs/gr-qc/0506126} {arXiv:gr-qc/0506126} \BibitemShut {NoStop}%
\bibitem [{\citenamefont {Fan}\ and\ \citenamefont {Wang}(2016)}]{Fan:2016hvf}%
  \BibitemOpen
  \bibfield  {author} {\bibinfo {author} {\bibfnamefont {Zhong-Ying}\ \bibnamefont {Fan}}\ and\ \bibinfo {author} {\bibfnamefont {Xiaobao}\ \bibnamefont {Wang}},\ }\bibfield  {title} {\enquote {\bibinfo {title} {{Construction of Regular Black Holes in General Relativity}},}\ }\href {\doibase 10.1103/PhysRevD.94.124027} {\bibfield  {journal} {\bibinfo  {journal} {Phys. Rev. D}\ }\textbf {\bibinfo {volume} {94}},\ \bibinfo {pages} {124027} (\bibinfo {year} {2016})},\ \Eprint {http://arxiv.org/abs/1610.02636} {arXiv:1610.02636 [gr-qc]} \BibitemShut {NoStop}%
\bibitem [{\citenamefont {Dymnikova}(1992)}]{Dymnikova:1992ux}%
  \BibitemOpen
  \bibfield  {author} {\bibinfo {author} {\bibfnamefont {I.}~\bibnamefont {Dymnikova}},\ }\bibfield  {title} {\enquote {\bibinfo {title} {{Vacuum nonsingular black hole}},}\ }\href {\doibase 10.1007/BF00760226} {\bibfield  {journal} {\bibinfo  {journal} {Gen. Rel. Grav.}\ }\textbf {\bibinfo {volume} {24}},\ \bibinfo {pages} {235--242} (\bibinfo {year} {1992})}\BibitemShut {NoStop}%
\bibitem [{\citenamefont {Green}(2024)}]{Green:2024bam}%
  \BibitemOpen
  \bibfield  {author} {\bibinfo {author} {\bibfnamefont {Anne~M.}\ \bibnamefont {Green}},\ }\bibfield  {title} {\enquote {\bibinfo {title} {{Primordial black holes as a dark matter candidate - a brief overview}},}\ }\href {\doibase 10.1016/j.nuclphysb.2024.116494} {\bibfield  {journal} {\bibinfo  {journal} {Nucl. Phys. B}\ }\textbf {\bibinfo {volume} {1003}},\ \bibinfo {pages} {116494} (\bibinfo {year} {2024})},\ \Eprint {http://arxiv.org/abs/2402.15211} {arXiv:2402.15211 [astro-ph.CO]} \BibitemShut {NoStop}%
\bibitem [{\citenamefont {Carr}(1975)}]{Carr:1975qj}%
  \BibitemOpen
  \bibfield  {author} {\bibinfo {author} {\bibfnamefont {Bernard~J.}\ \bibnamefont {Carr}},\ }\bibfield  {title} {\enquote {\bibinfo {title} {{The Primordial black hole mass spectrum}},}\ }\href {\doibase 10.1086/153853} {\bibfield  {journal} {\bibinfo  {journal} {Astrophys. J.}\ }\textbf {\bibinfo {volume} {201}},\ \bibinfo {pages} {1--19} (\bibinfo {year} {1975})}\BibitemShut {NoStop}%
\bibitem [{\citenamefont {MacGibbon}\ and\ \citenamefont {Carr}(1991)}]{MacGibbon:1991vc}%
  \BibitemOpen
  \bibfield  {author} {\bibinfo {author} {\bibfnamefont {Jane~H.}\ \bibnamefont {MacGibbon}}\ and\ \bibinfo {author} {\bibfnamefont {Bernard~J.}\ \bibnamefont {Carr}},\ }\bibfield  {title} {\enquote {\bibinfo {title} {{Cosmic rays from primordial black holes}},}\ }\href {\doibase 10.1086/169909} {\bibfield  {journal} {\bibinfo  {journal} {Astrophys. J.}\ }\textbf {\bibinfo {volume} {371}},\ \bibinfo {pages} {447--469} (\bibinfo {year} {1991})}\BibitemShut {NoStop}%
\bibitem [{\citenamefont {Carr}\ \emph {et~al.}(2010)\citenamefont {Carr}, \citenamefont {Kohri}, \citenamefont {Sendouda},\ and\ \citenamefont {Yokoyama}}]{Carr:2009jm}%
  \BibitemOpen
  \bibfield  {author} {\bibinfo {author} {\bibfnamefont {B.~J.}\ \bibnamefont {Carr}}, \bibinfo {author} {\bibfnamefont {Kazunori}\ \bibnamefont {Kohri}}, \bibinfo {author} {\bibfnamefont {Yuuiti}\ \bibnamefont {Sendouda}}, \ and\ \bibinfo {author} {\bibfnamefont {Jun'ichi}\ \bibnamefont {Yokoyama}},\ }\bibfield  {title} {\enquote {\bibinfo {title} {{New cosmological constraints on primordial black holes}},}\ }\href {\doibase 10.1103/PhysRevD.81.104019} {\bibfield  {journal} {\bibinfo  {journal} {Phys. Rev. D}\ }\textbf {\bibinfo {volume} {81}},\ \bibinfo {pages} {104019} (\bibinfo {year} {2010})},\ \Eprint {http://arxiv.org/abs/0912.5297} {arXiv:0912.5297 [astro-ph.CO]} \BibitemShut {NoStop}%
\bibitem [{\citenamefont {Laha}(2019)}]{Laha:2019ssq}%
  \BibitemOpen
  \bibfield  {author} {\bibinfo {author} {\bibfnamefont {Ranjan}\ \bibnamefont {Laha}},\ }\bibfield  {title} {\enquote {\bibinfo {title} {{Primordial Black Holes as a Dark Matter Candidate Are Severely Constrained by the Galactic Center 511 keV $\gamma$ -Ray Line}},}\ }\href {\doibase 10.1103/PhysRevLett.123.251101} {\bibfield  {journal} {\bibinfo  {journal} {Phys. Rev. Lett.}\ }\textbf {\bibinfo {volume} {123}},\ \bibinfo {pages} {251101} (\bibinfo {year} {2019})},\ \Eprint {http://arxiv.org/abs/1906.09994} {arXiv:1906.09994 [astro-ph.HE]} \BibitemShut {NoStop}%
\bibitem [{\citenamefont {Boudaud}\ and\ \citenamefont {Cirelli}(2019)}]{Boudaud:2018hqb}%
  \BibitemOpen
  \bibfield  {author} {\bibinfo {author} {\bibfnamefont {Mathieu}\ \bibnamefont {Boudaud}}\ and\ \bibinfo {author} {\bibfnamefont {Marco}\ \bibnamefont {Cirelli}},\ }\bibfield  {title} {\enquote {\bibinfo {title} {{Voyager 1 $e^\pm$ Further Constrain Primordial Black Holes as Dark Matter}},}\ }\href {\doibase 10.1103/PhysRevLett.122.041104} {\bibfield  {journal} {\bibinfo  {journal} {Phys. Rev. Lett.}\ }\textbf {\bibinfo {volume} {122}},\ \bibinfo {pages} {041104} (\bibinfo {year} {2019})},\ \Eprint {http://arxiv.org/abs/1807.03075} {arXiv:1807.03075 [astro-ph.HE]} \BibitemShut {NoStop}%
\bibitem [{\citenamefont {Easson}\ and\ \citenamefont {Brandenberger}(2001)}]{Easson:2001qf}%
  \BibitemOpen
  \bibfield  {author} {\bibinfo {author} {\bibfnamefont {Damien~A.}\ \bibnamefont {Easson}}\ and\ \bibinfo {author} {\bibfnamefont {Robert~H.}\ \bibnamefont {Brandenberger}},\ }\bibfield  {title} {\enquote {\bibinfo {title} {{Universe generation from black hole interiors}},}\ }\href {\doibase 10.1088/1126-6708/2001/06/024} {\bibfield  {journal} {\bibinfo  {journal} {JHEP}\ }\textbf {\bibinfo {volume} {06}},\ \bibinfo {pages} {024} (\bibinfo {year} {2001})},\ \Eprint {http://arxiv.org/abs/hep-th/0103019} {arXiv:hep-th/0103019} \BibitemShut {NoStop}%
\bibitem [{\citenamefont {Hawking}\ and\ \citenamefont {Penrose}(1970)}]{Hawking:1970zqf}%
  \BibitemOpen
  \bibfield  {author} {\bibinfo {author} {\bibfnamefont {S.~W.}\ \bibnamefont {Hawking}}\ and\ \bibinfo {author} {\bibfnamefont {R.}~\bibnamefont {Penrose}},\ }\bibfield  {title} {\enquote {\bibinfo {title} {{The Singularities of gravitational collapse and cosmology}},}\ }\href {\doibase 10.1098/rspa.1970.0021} {\bibfield  {journal} {\bibinfo  {journal} {Proc. Roy. Soc. Lond. A}\ }\textbf {\bibinfo {volume} {314}},\ \bibinfo {pages} {529--548} (\bibinfo {year} {1970})}\BibitemShut {NoStop}%
\bibitem [{\citenamefont {Zaslavskii}(2010)}]{ZASLAVSKII2010278}%
  \BibitemOpen
  \bibfield  {author} {\bibinfo {author} {\bibfnamefont {O.B.}\ \bibnamefont {Zaslavskii}},\ }\bibfield  {title} {\enquote {\bibinfo {title} {Regular black holes and energy conditions},}\ }\href {\doibase https://doi.org/10.1016/j.physletb.2010.04.031} {\bibfield  {journal} {\bibinfo  {journal} {Physics Letters B}\ }\textbf {\bibinfo {volume} {688}},\ \bibinfo {pages} {278--280} (\bibinfo {year} {2010})}\BibitemShut {NoStop}%
\bibitem [{\citenamefont {Maeda}(2022)}]{Maeda:2021jdc}%
  \BibitemOpen
  \bibfield  {author} {\bibinfo {author} {\bibfnamefont {Hideki}\ \bibnamefont {Maeda}},\ }\bibfield  {title} {\enquote {\bibinfo {title} {{Quest for realistic non-singular black-hole geometries: regular-center type}},}\ }\href {\doibase 10.1007/JHEP11(2022)108} {\bibfield  {journal} {\bibinfo  {journal} {JHEP}\ }\textbf {\bibinfo {volume} {11}},\ \bibinfo {pages} {108} (\bibinfo {year} {2022})},\ \Eprint {http://arxiv.org/abs/2107.04791} {arXiv:2107.04791 [gr-qc]} \BibitemShut {NoStop}%
\bibitem [{\citenamefont {Birkhoff}(1923)}]{Birkhoff1923}%
  \BibitemOpen
  \bibfield  {author} {\bibinfo {author} {\bibfnamefont {G.~D.}\ \bibnamefont {Birkhoff}},\ }\href@noop {} {\bibfield  {journal} {\bibinfo  {journal} {Relativity and Modern Physics, Harvard University Press,}\ ,\ \bibinfo {pages} {p. 253}} (\bibinfo {year} {1923})}\BibitemShut {NoStop}%
\bibitem [{\citenamefont {Easson}\ and\ \citenamefont {Pezzelle}(2024)}]{Easson:2023ytf}%
  \BibitemOpen
  \bibfield  {author} {\bibinfo {author} {\bibfnamefont {Damien~A.}\ \bibnamefont {Easson}}\ and\ \bibinfo {author} {\bibfnamefont {Max~W.}\ \bibnamefont {Pezzelle}},\ }\bibfield  {title} {\enquote {\bibinfo {title} {{Kleinian black holes}},}\ }\href {\doibase 10.1103/PhysRevD.109.044007} {\bibfield  {journal} {\bibinfo  {journal} {Phys. Rev. D}\ }\textbf {\bibinfo {volume} {109}},\ \bibinfo {pages} {044007} (\bibinfo {year} {2024})},\ \Eprint {http://arxiv.org/abs/2312.00879} {arXiv:2312.00879 [hep-th]} \BibitemShut {NoStop}%
\bibitem [{\citenamefont {Chatterjee}\ \emph {et~al.}(2013)\citenamefont {Chatterjee}, \citenamefont {Easson},\ and\ \citenamefont {Parikh}}]{Chatterjee:2012zh}%
  \BibitemOpen
  \bibfield  {author} {\bibinfo {author} {\bibfnamefont {Saugata}\ \bibnamefont {Chatterjee}}, \bibinfo {author} {\bibfnamefont {Damien~A.}\ \bibnamefont {Easson}}, \ and\ \bibinfo {author} {\bibfnamefont {Maulik}\ \bibnamefont {Parikh}},\ }\bibfield  {title} {\enquote {\bibinfo {title} {{Energy conditions in the Jordan frame}},}\ }\href {\doibase 10.1088/0264-9381/30/23/235031} {\bibfield  {journal} {\bibinfo  {journal} {Class. Quant. Grav.}\ }\textbf {\bibinfo {volume} {30}},\ \bibinfo {pages} {235031} (\bibinfo {year} {2013})},\ \Eprint {http://arxiv.org/abs/1212.6430} {arXiv:1212.6430 [gr-qc]} \BibitemShut {NoStop}%
\bibitem [{\citenamefont {Penrose}(1965)}]{Penrose:1964wq}%
  \BibitemOpen
  \bibfield  {author} {\bibinfo {author} {\bibfnamefont {Roger}\ \bibnamefont {Penrose}},\ }\bibfield  {title} {\enquote {\bibinfo {title} {{Gravitational collapse and space-time singularities}},}\ }\href {\doibase 10.1103/PhysRevLett.14.57} {\bibfield  {journal} {\bibinfo  {journal} {Phys. Rev. Lett.}\ }\textbf {\bibinfo {volume} {14}},\ \bibinfo {pages} {57--59} (\bibinfo {year} {1965})}\BibitemShut {NoStop}%
\bibitem [{\citenamefont {Hawking}(1966{\natexlab{a}})}]{Hawking:1966sx}%
  \BibitemOpen
  \bibfield  {author} {\bibinfo {author} {\bibfnamefont {Stephen}\ \bibnamefont {Hawking}},\ }\bibfield  {title} {\enquote {\bibinfo {title} {{The Occurrence of singularities in cosmology}},}\ }\href {\doibase 10.1098/rspa.1966.0221} {\bibfield  {journal} {\bibinfo  {journal} {Proc. Roy. Soc. Lond. A}\ }\textbf {\bibinfo {volume} {294}},\ \bibinfo {pages} {511--521} (\bibinfo {year} {1966}{\natexlab{a}})}\BibitemShut {NoStop}%
\bibitem [{\citenamefont {Hawking}(1966{\natexlab{b}})}]{Hawking:1966jv}%
  \BibitemOpen
  \bibfield  {author} {\bibinfo {author} {\bibfnamefont {Stephen}\ \bibnamefont {Hawking}},\ }\bibfield  {title} {\enquote {\bibinfo {title} {{The Occurrence of singularities in cosmology. II}},}\ }\href {\doibase 10.1098/rspa.1966.0255} {\bibfield  {journal} {\bibinfo  {journal} {Proc. Roy. Soc. Lond. A}\ }\textbf {\bibinfo {volume} {295}},\ \bibinfo {pages} {490--493} (\bibinfo {year} {1966}{\natexlab{b}})}\BibitemShut {NoStop}%
\bibitem [{\citenamefont {Hawking}(1967)}]{Hawking:1967ju}%
  \BibitemOpen
  \bibfield  {author} {\bibinfo {author} {\bibfnamefont {Stephen}\ \bibnamefont {Hawking}},\ }\bibfield  {title} {\enquote {\bibinfo {title} {{The occurrence of singularities in cosmology. III. Causality and singularities}},}\ }\href {\doibase 10.1098/rspa.1967.0164} {\bibfield  {journal} {\bibinfo  {journal} {Proc. Roy. Soc. Lond. A}\ }\textbf {\bibinfo {volume} {300}},\ \bibinfo {pages} {187--201} (\bibinfo {year} {1967})}\BibitemShut {NoStop}%
\bibitem [{\citenamefont {Lesnefsky}\ \emph {et~al.}(2023)\citenamefont {Lesnefsky}, \citenamefont {Easson},\ and\ \citenamefont {Davies}}]{Lesnefsky:2022fen}%
  \BibitemOpen
  \bibfield  {author} {\bibinfo {author} {\bibfnamefont {J.~E.}\ \bibnamefont {Lesnefsky}}, \bibinfo {author} {\bibfnamefont {D.~A.}\ \bibnamefont {Easson}}, \ and\ \bibinfo {author} {\bibfnamefont {P.~C.~W.}\ \bibnamefont {Davies}},\ }\bibfield  {title} {\enquote {\bibinfo {title} {{Past-completeness of inflationary spacetimes}},}\ }\href {\doibase 10.1103/PhysRevD.107.044024} {\bibfield  {journal} {\bibinfo  {journal} {Phys. Rev. D}\ }\textbf {\bibinfo {volume} {107}},\ \bibinfo {pages} {044024} (\bibinfo {year} {2023})},\ \Eprint {http://arxiv.org/abs/2207.00955} {arXiv:2207.00955 [gr-qc]} \BibitemShut {NoStop}%
\bibitem [{\citenamefont {Easson}\ and\ \citenamefont {Lesnefsky}(2024{\natexlab{a}})}]{Easson:2024uxe}%
  \BibitemOpen
  \bibfield  {author} {\bibinfo {author} {\bibfnamefont {Damien~A.}\ \bibnamefont {Easson}}\ and\ \bibinfo {author} {\bibfnamefont {Joseph~E.}\ \bibnamefont {Lesnefsky}},\ }\bibfield  {title} {\enquote {\bibinfo {title} {{Inflationary resolution of the initial singularity}},}\ }\href@noop {} {\  (\bibinfo {year} {2024}{\natexlab{a}})},\ \Eprint {http://arxiv.org/abs/2402.13031} {arXiv:2402.13031 [hep-th]} \BibitemShut {NoStop}%
\bibitem [{\citenamefont {Easson}\ and\ \citenamefont {Lesnefsky}(2024{\natexlab{b}})}]{Easson:2024fzn}%
  \BibitemOpen
  \bibfield  {author} {\bibinfo {author} {\bibfnamefont {Damien~A.}\ \bibnamefont {Easson}}\ and\ \bibinfo {author} {\bibfnamefont {Joseph~E.}\ \bibnamefont {Lesnefsky}},\ }\bibfield  {title} {\enquote {\bibinfo {title} {{Eternal Universes}},}\ }\href@noop {} {\  (\bibinfo {year} {2024}{\natexlab{b}})},\ \Eprint {http://arxiv.org/abs/2404.03016} {arXiv:2404.03016 [hep-th]} \BibitemShut {NoStop}%
\bibitem [{\citenamefont {S.W.Hawking}\ and\ \citenamefont {G.F.R.Ellis}(1973)}]{S.W.Hawking1973}%
  \BibitemOpen
  \bibfield  {author} {\bibinfo {author} {\bibnamefont {S.W.Hawking}}\ and\ \bibinfo {author} {\bibnamefont {G.F.R.Ellis}},\ }\href@noop {} {\emph {\bibinfo {title} {{The large scale structure}}}}\ (\bibinfo {year} {1973})\BibitemShut {NoStop}%
\bibitem [{\citenamefont {Easson}\ \emph {et~al.}(2020)\citenamefont {Easson}, \citenamefont {Keeler},\ and\ \citenamefont {Manton}}]{Easson:2020esh}%
  \BibitemOpen
  \bibfield  {author} {\bibinfo {author} {\bibfnamefont {Damien~A.}\ \bibnamefont {Easson}}, \bibinfo {author} {\bibfnamefont {Cynthia}\ \bibnamefont {Keeler}}, \ and\ \bibinfo {author} {\bibfnamefont {Tucker}\ \bibnamefont {Manton}},\ }\bibfield  {title} {\enquote {\bibinfo {title} {{Classical double copy of nonsingular black holes}},}\ }\href {\doibase 10.1103/PhysRevD.102.086015} {\bibfield  {journal} {\bibinfo  {journal} {Phys. Rev. D}\ }\textbf {\bibinfo {volume} {102}},\ \bibinfo {pages} {086015} (\bibinfo {year} {2020})},\ \Eprint {http://arxiv.org/abs/2007.16186} {arXiv:2007.16186 [gr-qc]} \BibitemShut {NoStop}%
\bibitem [{\citenamefont {Ayon-Beato}\ and\ \citenamefont {Garcia}(1998)}]{Ayon-Beato:1998hmi}%
  \BibitemOpen
  \bibfield  {author} {\bibinfo {author} {\bibfnamefont {Eloy}\ \bibnamefont {Ayon-Beato}}\ and\ \bibinfo {author} {\bibfnamefont {Alberto}\ \bibnamefont {Garcia}},\ }\bibfield  {title} {\enquote {\bibinfo {title} {{Regular black hole in general relativity coupled to nonlinear electrodynamics}},}\ }\href {\doibase 10.1103/PhysRevLett.80.5056} {\bibfield  {journal} {\bibinfo  {journal} {Phys. Rev. Lett.}\ }\textbf {\bibinfo {volume} {80}},\ \bibinfo {pages} {5056--5059} (\bibinfo {year} {1998})},\ \Eprint {http://arxiv.org/abs/gr-qc/9911046} {arXiv:gr-qc/9911046} \BibitemShut {NoStop}%
\bibitem [{\citenamefont {Ayon-Beato}\ and\ \citenamefont {Garcia}(2000)}]{Ayon-Beato:2000mjt}%
  \BibitemOpen
  \bibfield  {author} {\bibinfo {author} {\bibfnamefont {Eloy}\ \bibnamefont {Ayon-Beato}}\ and\ \bibinfo {author} {\bibfnamefont {Alberto}\ \bibnamefont {Garcia}},\ }\bibfield  {title} {\enquote {\bibinfo {title} {{The Bardeen model as a nonlinear magnetic monopole}},}\ }\href {\doibase 10.1016/S0370-2693(00)01125-4} {\bibfield  {journal} {\bibinfo  {journal} {Phys. Lett. B}\ }\textbf {\bibinfo {volume} {493}},\ \bibinfo {pages} {149--152} (\bibinfo {year} {2000})},\ \Eprint {http://arxiv.org/abs/gr-qc/0009077} {arXiv:gr-qc/0009077} \BibitemShut {NoStop}%
\bibitem [{\citenamefont {Gaete}\ \emph {et~al.}(2022)\citenamefont {Gaete}, \citenamefont {Jusufi},\ and\ \citenamefont {Nicolini}}]{Gaete:2022ukm}%
  \BibitemOpen
  \bibfield  {author} {\bibinfo {author} {\bibfnamefont {Patricio}\ \bibnamefont {Gaete}}, \bibinfo {author} {\bibfnamefont {Kimet}\ \bibnamefont {Jusufi}}, \ and\ \bibinfo {author} {\bibfnamefont {Piero}\ \bibnamefont {Nicolini}},\ }\bibfield  {title} {\enquote {\bibinfo {title} {{Charged black holes from T-duality}},}\ }\href {\doibase 10.1016/j.physletb.2022.137546} {\bibfield  {journal} {\bibinfo  {journal} {Phys. Lett. B}\ }\textbf {\bibinfo {volume} {835}},\ \bibinfo {pages} {137546} (\bibinfo {year} {2022})},\ \Eprint {http://arxiv.org/abs/2205.15441} {arXiv:2205.15441 [hep-th]} \BibitemShut {NoStop}%
\bibitem [{\citenamefont {Bueno}\ \emph {et~al.}(2024{\natexlab{a}})\citenamefont {Bueno}, \citenamefont {Cano}, \citenamefont {Hennigar},\ and\ \citenamefont {Murcia}}]{Bueno:2024eig}%
  \BibitemOpen
  \bibfield  {author} {\bibinfo {author} {\bibfnamefont {Pablo}\ \bibnamefont {Bueno}}, \bibinfo {author} {\bibfnamefont {Pablo~A.}\ \bibnamefont {Cano}}, \bibinfo {author} {\bibfnamefont {Robie~A.}\ \bibnamefont {Hennigar}}, \ and\ \bibinfo {author} {\bibfnamefont {\'Angel~J.}\ \bibnamefont {Murcia}},\ }\bibfield  {title} {\enquote {\bibinfo {title} {{Dynamical Formation of Regular Black Holes}},}\ }\href@noop {} {\  (\bibinfo {year} {2024}{\natexlab{a}})},\ \Eprint {http://arxiv.org/abs/2412.02742} {arXiv:2412.02742 [gr-qc]} \BibitemShut {NoStop}%
\bibitem [{\citenamefont {Bueno}\ \emph {et~al.}(2024{\natexlab{b}})\citenamefont {Bueno}, \citenamefont {Cano}, \citenamefont {Hennigar},\ and\ \citenamefont {Murcia}}]{Bueno:2024zsx}%
  \BibitemOpen
  \bibfield  {author} {\bibinfo {author} {\bibfnamefont {Pablo}\ \bibnamefont {Bueno}}, \bibinfo {author} {\bibfnamefont {Pablo~A.}\ \bibnamefont {Cano}}, \bibinfo {author} {\bibfnamefont {Robie~A.}\ \bibnamefont {Hennigar}}, \ and\ \bibinfo {author} {\bibfnamefont {\'Angel~J.}\ \bibnamefont {Murcia}},\ }\bibfield  {title} {\enquote {\bibinfo {title} {{Regular black holes from thin-shell collapse}},}\ }\href@noop {} {\  (\bibinfo {year} {2024}{\natexlab{b}})},\ \Eprint {http://arxiv.org/abs/2412.02740} {arXiv:2412.02740 [gr-qc]} \BibitemShut {NoStop}%
\bibitem [{\citenamefont {Yan}\ \emph {et~al.}(2020)\citenamefont {Yan}, \citenamefont {Li}, \citenamefont {Xue}, \citenamefont {Ren}, \citenamefont {Cai}, \citenamefont {Easson}, \citenamefont {Yuan},\ and\ \citenamefont {Zhao}}]{Yan:2019hxx}%
  \BibitemOpen
  \bibfield  {author} {\bibinfo {author} {\bibfnamefont {Sheng-Feng}\ \bibnamefont {Yan}}, \bibinfo {author} {\bibfnamefont {Chunlong}\ \bibnamefont {Li}}, \bibinfo {author} {\bibfnamefont {Lingqin}\ \bibnamefont {Xue}}, \bibinfo {author} {\bibfnamefont {Xin}\ \bibnamefont {Ren}}, \bibinfo {author} {\bibfnamefont {Yi-Fu}\ \bibnamefont {Cai}}, \bibinfo {author} {\bibfnamefont {Damien~A.}\ \bibnamefont {Easson}}, \bibinfo {author} {\bibfnamefont {Ye-Fei}\ \bibnamefont {Yuan}}, \ and\ \bibinfo {author} {\bibfnamefont {Hongsheng}\ \bibnamefont {Zhao}},\ }\bibfield  {title} {\enquote {\bibinfo {title} {{Testing the equivalence principle via the shadow of black holes}},}\ }\href {\doibase 10.1103/PhysRevResearch.2.023164} {\bibfield  {journal} {\bibinfo  {journal} {Phys. Rev. Res.}\ }\textbf {\bibinfo {volume} {2}},\ \bibinfo {pages} {023164} (\bibinfo {year} {2020})},\ \Eprint {http://arxiv.org/abs/1912.12629} {arXiv:1912.12629 [astro-ph.CO]} \BibitemShut {NoStop}%
\bibitem [{\citenamefont {Balart}\ \emph {et~al.}(2025)\citenamefont {Balart}, \citenamefont {Panotopoulos},\ and\ \citenamefont {Rinc\'on}}]{Balart:2024rtj}%
  \BibitemOpen
  \bibfield  {author} {\bibinfo {author} {\bibfnamefont {Leonardo}\ \bibnamefont {Balart}}, \bibinfo {author} {\bibfnamefont {Grigoris}\ \bibnamefont {Panotopoulos}}, \ and\ \bibinfo {author} {\bibfnamefont {\'Angel}\ \bibnamefont {Rinc\'on}},\ }\bibfield  {title} {\enquote {\bibinfo {title} {{On new regular charged black hole solutions: Limiting Curvature Condition, Quasinormal modes and Shadows}},}\ }\href {\doibase 10.1016/j.aop.2024.169865} {\bibfield  {journal} {\bibinfo  {journal} {Annals Phys.}\ }\textbf {\bibinfo {volume} {473}},\ \bibinfo {pages} {169865} (\bibinfo {year} {2025})},\ \Eprint {http://arxiv.org/abs/2412.00550} {arXiv:2412.00550 [gr-qc]} \BibitemShut {NoStop}%
\bibitem [{\citenamefont {Vertogradov}\ \emph {et~al.}(2025)\citenamefont {Vertogradov}, \citenamefont {\"Ovg\"un},\ and\ \citenamefont {Shatov}}]{Vertogradov:2025snh}%
  \BibitemOpen
  \bibfield  {author} {\bibinfo {author} {\bibfnamefont {Vitalii}\ \bibnamefont {Vertogradov}}, \bibinfo {author} {\bibfnamefont {Ali}\ \bibnamefont {\"Ovg\"un}}, \ and\ \bibinfo {author} {\bibfnamefont {Daniil}\ \bibnamefont {Shatov}},\ }\bibfield  {title} {\enquote {\bibinfo {title} {{Formation of regular black hole from baryonic matter}},}\ }\href@noop {} {\  (\bibinfo {year} {2025})},\ \Eprint {http://arxiv.org/abs/2502.00521} {arXiv:2502.00521 [gr-qc]} \BibitemShut {NoStop}%
\end{thebibliography}%
\end{document}